\documentclass[11pt]{article}
\def \d {{\rm d}}
\def \r {{\rm r}}
\def \t {{\rm t}}
\def \vv {{\rm v}}
\def \H {{\rm H}}
\def \AA {{\rm A}}
\def \BB {{\rm B}}
\newcommand{\A}{{\cal A}}
\newcommand{\R}{{\cal R}}
\newcommand{\I}{{\cal I}}
\def \U {{\cal U}}
\def \V {{\cal V}}

\newcommand{\e}{{\bf e}}
\newcommand{\pul}{\frac{1}{2}}
\newcommand{\la}{\frac{\Lambda}{6}}

\begin{document}

\title{Exact impulsive gravitational waves in \\
       spacetimes of constant curvature}

\author{J. Podolsk\'y\thanks{E--mail: {\tt podolsky@mbox.troja.mff.cuni.cz}}
\\
\\ Institute of Theoretical Physics, Charles University in Prague,\\
V Hole\v{s}ovi\v{c}k\'ach 2, 180\ 00 Prague 8, Czech Republic.\\}

\date{January 7, 2002}

\maketitle

\begin{abstract}
Exact solutions exist which describe impulsive gravitational waves
propagating in Minkowski, de~Sitter, or anti-de~Sitter universes. These
may be either nonexpanding or expanding. Both cases in each background
are reviewed here from a unified point of view. All the main methods for
their construction are described systematically: the Penrose ``cut
and paste'' method, explicit construction of continuous coordinates,
distributional limits of sandwich waves, embedding from higher
dimensions, and boosts of sources or limits of infinite accelerations.

Attention is concentrated on the most interesting specific solutions.
In particular, the nonexpanding impulsive waves that are generated by
null multipole particles are described. These generalize the
well-known  Aichelburg--Sexl  and Hotta--Tanaka monopole solutions.
Also described are the expanding spherical impulses that are
generated by snapping and
colliding strings. Geodesics and some other properties of impulsive
wave spacetimes are also summarized.
\end{abstract}

\section{Outline and personal notes}

In recent years, various aspects of exact solutions of  Einstein's equations
which describe impulsive  waves in spaces of constant curvature have been
thoroughly investigated. However, these studies are mostly scattered in a
great number of particular articles published in different journals. The first
purpose of this essay is to collect and unify the principal results. We attempt to
provide a brief but comprehensive and self-contained review of all impulsive
gravitational waves which may propagate in Minkowski, de~Sitter, or anti-de~Sitter
universe. We hope that such a summary may be useful for further studies in
classical relativity and also  in quantum gravity or string theory.

Nevertheless, the main motivation for our work was to pay a tribute to Professor
J.~Bi\v c\'ak on the occasion of his 60th birthday.  Ji\v r\'\i \ Bi\v c\'ak's
lifelong devotion to the investigation of exact radiative spacetimes
is famous. He has contributed to various specific topics, including those
closely related to impulsive waves, which we survey here.
But Professor Bi\v c\'ak not only wrote many contributions which will
be mentioned below in the appropriate context. During the past decades he has
also been an inexhaustible fountain-head of inspiration and a continuous source
of encouragement --- and not only to members of the Prague relativity group.

Even more personally, it
was Ji\v r\'\i \ Bi\v c\'ak --- my academic teacher --- who in 1985 suggested the
topic of my diploma thesis  ``Gravitational radiation in
cosmology''. Later he carefully supervised my doctoral thesis ``On exact radiative
space-times with cosmological constant''. During all these years I have been learning
fundamental issues concerning the theory of  gravitational radiation from him.
In particular, the geometrical and physical  investigation of the Kundt and the Robinson--Trautman
classes of exact gravitational waves later proved  to be crucial for my understanding
of possible impulses in spaces of constant curvature. Moreover,
almost all of my recent works on impulsive solutions have emerged from a fruitful
collaboration with Professor J.~B.~Griffiths from Loughborough University.
And again, it was Ji\v r\'\i \ Bi\v c\'ak who established our
mutual contacts and has been promoting and encouraging these during the last years $\dots$
It is thus only natural to dedicate this summarizing essay to him on the occasion of
the exceptional anniversary of his birthday.

The present work is organized as follows. We shall first briefly
describe the background spaces of constant curvature, present all
(nontwisting) type $N$ exact vacuum solutions with a cosmological constant, and
``heuristically'' introduce gravitational impulses as limits of
corresponding sandwich waves. In the main section~3, we shall
describe  in detail the methods of construction of these impulsive waves
--- the Penrose ``cut and paste'' method, explicit construction of continuous
coordinates, distributional limits of sandwich waves,
geometrical embedding from higher dimensions, and boosts or limits of infinite
acceleration of particular solutions. Using all these methods, both
nonexpanding and expanding impulsive waves are constructed, as
summarized at the end of  section~3. The physically important particular
solutions are presented in section~4, namely
nonexpanding impulses generated by  multipole particles, and
expanding impulses generated by snapping or colliding strings.
Finally, some other properties are also discussed, in particular the behaviour
of geodesics.

\section{Exposition}

\subsection{Spaces of constant curvature}

The spaces of constant curvature ---  Minkowski, de Sitter, and anti-de Sitter
spacetimes --- which form  backgrounds for the impulsive solutions investigated
here, are the simplest exact solutions of Einstein's vacuum field
equations with vanishing, positive, and negative  cosmological constant $\Lambda$,
respectively. Yet, these are extremely important in contemporary theoretical
physics. As Ji\v r\'\i \ Bi\v c\'ak writes in his recent substantial
essay \cite{JiBi2000} in honour of J\"urgen Ehlers:
``These solutions have played a crucial
role in many issues in general relativity and cosmology, and most
recently, they have become important prerequisites on the stage
of the theoretical physics of the `new age', including string
theory and string cosmology''.
The de Sitter universe is the basic model of early quasi-exponential
inflationary phase of expansion of the universe \cite{Guth}.
Moreover, according to the ``cosmic no-hair'' conjecture \cite{Maeda},
it is locally the asymptotic state of many cosmological models
with a positive $\Lambda$.
Recently, the anti-de Sitter spacetime has also become a subject
of intensive studies thanks to the Maldacena's conjecture \cite{MDC}
which relates string theory in asymptotically anti-de Sitter space to a
non-gravitational conformal field theory on the boundary at
spatial infinity.

The spacetimes of  constant curvature are highly symmetric: they admit ten
Killing vector fields which correspond to the maximal number of
isometries possible in a 4-dimensional spacetime. Also, they are conformally
flat and can thus be understood as specific portions
of the Einstein static universe. Their global conformal structures
are however different for different signs of $\Lambda$ (see below).

It is well-known that the de~Sitter and anti-de~Sitter spacetimes can be
represented as the 4-dimensional hyperboloid
\begin{equation}
 -{Z_0}^2+{Z_1}^2+{Z_2}^2+{Z_3}^2+\epsilon{Z_4}^2=\epsilon a^2\ , \qquad
a=\sqrt{3 / |\Lambda|}\ ,  \label{hyp}
\end{equation}
 in the flat 5-dimensional space with metric
\begin{equation}
\d s_0^2=-\d{Z_0}^2+\d{Z_1}^2+\d{Z_2}^2+\d{Z_3}^2+\epsilon\d{Z_4}^2\ ,\label{ds0}
\end{equation}
where $\epsilon=1$ for $\Lambda>0$ and $\epsilon=-1$ for $\Lambda<0$.
Obviously, the de Sitter spacetime has the topology $R^1\times S^3$ ($R^1$
corresponding to the time), whereas the topology of the anti-de Sitter spacetime
is $S^1\times R^3$.

Various parameterizations of the hyperboloid (\ref{hyp})
are known which introduce suitable coordinates for the de~Sitter
or the anti-de~Sitter spacetime. For example, the expressions
 \begin{eqnarray}
 Z_0 &=& {\textstyle{1\over\sqrt2}(\U+\V)/
\left[1+{1\over6}\Lambda(\eta\bar\eta-\U\V)\right]}\ , \nonumber\\
 Z_1 &=& {\textstyle{1\over\sqrt2}(\U-\V)/
\left[1+{1\over6}\Lambda(\eta\bar\eta-\U\V)\right]}\ , \nonumber\\
 Z_2 &=& {\textstyle{1\over\sqrt2}(\eta+\bar\eta)/
\left[1+{1\over6}\Lambda(\eta\bar\eta-\U\V)\right]}\ , \label{Zcoords}\\
 Z_3 &=& {\textstyle{-\hbox{i}\over\sqrt2}(\eta-\bar\eta)/
\left[1+{1\over6}\Lambda(\eta\bar\eta-\U\V)\right]}\ , \nonumber\\
 Z_4 &=& a\,{\textstyle
\left[1-{1\over6}\Lambda(\eta\bar\eta-\U\V)\right]/
\left[1+{1\over6}\Lambda(\eta\bar\eta-\U\V)\right]}\ , \nonumber
 \end{eqnarray}
i.e. inversely $\U={\sqrt2 a}(Z_0+Z_1)/(Z_4+a)$,
$\V={\sqrt2 a}(Z_0-Z_1)/(Z_4+a)$, $\eta={\sqrt2
a}(Z_2+\hbox{i}Z_3)/(Z_4+a)$,
give the line element  for all spaces of constant curvature
in the  conformally flat unified form
 \begin{equation}
\d s_0^2= {2\,\d\eta\,\d\bar\eta -2\,\d\U\,\d\V
\over [\,1+{1\over6}\Lambda(\eta\bar\eta-\U\V)\,]^2}\ .
 \label{conf}
 \end{equation}
This is de~Sitter space when $\Lambda>0$, anti-de~Sitter space when
$\Lambda<0$, and Minkowski space when $\Lambda=0$.

However, the most natural coordinates which cover the whole de~Sitter hyperboloid
(\ref{hyp}) are
 \begin{eqnarray}
Z_0&=&a\sinh(\tau/a)\ , \nonumber\\
Z_1&=&a\cosh(\tau/a)\cos\chi\ , \nonumber\\
Z_2&=&a\cosh(\tau/a)\sin\chi\sin\theta\cos\phi\ , \label{globpar}\\
Z_3&=&a\cosh(\tau/a)\sin\chi\sin\theta\sin\phi\ , \nonumber\\
Z_4&=&a\cosh(\tau/a)\sin\chi\cos\theta\ .\nonumber
 \end{eqnarray}
With these  the line element (\ref{ds0}) becomes
 \begin{equation}
 \d s_0^2=-\d \tau^2 +a^2\cosh^2(\tau/a)\,[\,\d\chi^2
+\sin^2\chi\,(\d\theta^2+\sin^2\theta\,\d\phi^2)\,]\ . \label{glob}
 \end{equation}
 In this case, the privileged timelike observers have worldlines on
which the space coordinates $\chi$, $\theta$ and $\phi$ are constant. For this family,
the 3-spaces $\tau=$ const. (which represent the universe at this instant) have the geometry
of $S^3$ which contracts to a minimum size at $\tau=0$ and
then re-expand. The global conformal structure of the de~Sitter spacetime becomes
obvious if we introduce the conformal time $\eta_{_E}$ by
$\sin\eta_{_E}= 1/\cosh(\tau/a) $. The line element (\ref{glob}) then becomes
$\d s^2_0=\Omega^2\,\d s^2_{_E}$, where $\Omega=a/\sin\eta_{_E}$, and
$\d s^2_{_E}=-\d\eta_{_E}^2 +\d\chi^2 +\sin^2\chi\,(\d\theta^2+\sin^2\theta\,\d\phi^2)$
is the standard metric of the Einstein static universe.
Therefore, the corresponding Penrose conformal diagram covers the coordinate ranges
$\chi\in[0,\pi]$ and $\eta_{_E}\in(0,\pi)$. The boundaries of $\eta_{_E}$ (i.e.
$\tau=\pm\infty$) represent the conformal infinities $\cal J^\pm$. Obviously, these have a
spacelike character.

The natural global coordinate system for the anti-de~Sitter spacetime is obtained by
 \begin{eqnarray}
 Z_0&=&a\cosh (\r)\sin(\t/a)\ , \nonumber\\
 Z_4&=&a\cosh (\r)\cos(\t/a)\ , \nonumber\\
 Z_1&=&a\sinh (\r)\cos\theta\ , \label{ads}\\
 Z_2&=&a\sinh (\r)\sin\theta\cos\phi\ , \nonumber\\
 Z_3&=&a\sinh (\r)\sin\theta\sin\phi\ , \nonumber
 \end{eqnarray}
with which the line element (\ref{ds0}) becomes
 \begin{equation}
 \d s_0^2=-\cosh^2\!\r\,\,\d \t^2 +a^2\,[\,\d \r^2
+\sinh^2\!\r\,(\d\theta^2+\sin^2\theta\,\d\phi^2)\,]\ .
\end{equation}
Again, the privileged  timelike (non-geodesic) observers have
worldlines   $\r$, $\theta$ and $\phi$ constants, their proper times
are given by $\tau=\t\cosh(\r)$. However, the timelike slices $\t=$ const.
are 3-dimensional spaces of constant negative curvature. The conformal structure
is seen by introducing $\eta_{_E}=\t/a$ and $\tan\chi=\sinh(\r)$. Then
$\>\d s^2_0=\Omega^2\,\d s^2_{_E}\,$, where $\Omega=a/\cos\chi$, so
that the conformal diagram of the anti-de~Sitter spacetime is given by
the range $\chi\in[0,{\pi\over2})$, $\eta_{_E}$ arbitrary. Thus,
contrary to the Minkowski or de~Sitter universe, the conformal infinity $\cal J$
 corresponding to $\chi={\pi\over2}$ has a timelike character in the anti-de~Sitter
spacetime.

Fundamental geometrical properties of  the spaces of constant curvature have
been described in detail in \cite{Pen} and in more recent works and reviews,
see e.g.  \cite{Wa}. Note also that the Minkowski, de Sitter, and  anti-de Sitter
spacetimes are stable with respect to general, nonlinear (but ``weak'') vacuum
perturbations, as shown recently in \cite{CHK}, \cite{HF}.

\subsection{Exact type $N$ Einstein spaces}

Impulsive waves  can (at least from a physical point of view) most
naturally be understood as ``distributional limits'' of suitable families of
 waves with ``sandwich'' profiles, such that these have ever
``shorter duration'' but simultaneously become ``stronger'' in the limit.
Impulsive gravitational waves can be intuitively understood in exactly this way.
Incidentally, in our   contributions  \cite{[A1]}, \cite{[A2]} with  Ji\v r\'\i \ Bi\v c\'ak
we investigated  classes of exact radiative spacetimes, which may be used for the
above construction. It will thus be useful and convenient for
our later purposes to summarize the fundamental properties
of these  spacetimes here.

In particular, we studied all nontwisting Petrov-type {\it N} solutions of vacuum
Einstein field equations with a cosmological constant. For type {\it N\,}
spacetimes, the Debever--Penrose  vector field $k^\alpha$ is quadruple
and defines a privileged null congruence characterized by
expansion $\Theta = \pul k^\alpha _{\ ;\alpha} $,
twist $\omega = \sqrt{\pul k_{[\alpha ; \beta]}k^{\alpha ; \beta}}$,
and shear $\left|\sigma \right|= \sqrt{\pul k_{(\alpha;\beta)}k^{\alpha;\beta}
  - \Theta^2}$.
The Bianchi identities and the Kundt--Thompson theorem for
type {\it N\,} Einstein spaces give $\sigma=0$.
Considering the nontwisting case only, we are left with two classes:

\begin{itemize}

\item  the {\it Kundt} class of {\it nonexpanding} gravitational waves
($\Theta=0$), cf. \cite{Kundt}, \cite{ORR}, Ch.~27 in \cite{KSMH},

\item  the {\it Robinson--Trautman} class of {\it expanding} gravitational
waves ($\Theta\not=0$), cf. \cite{RT}, Ch.~24 in \cite{KSMH}.

\end{itemize}

\noindent
Hereafter, we denote the above Kundt class by $KN(\Lambda)$, and the
Robinson--Trautman class by $RTN(\Lambda)$.

In \cite{[A1]} we described these spacetimes in detail from
a geometrical point of view in a unified formalism, we presented
invariant subclasses, and found  relations between them.
All solutions of the Kundt class $KN(\Lambda)$ can be written in suitable
coordinates, first given in \cite{ORR}, as
\begin{equation}
\d s^2=2\,{1\over p^2}\,\d\xi\d\bar\xi - 2\ {q^2\over p^2}\,\d
u\d v+ {\rm F}\, \d u^2
\ ,\label{KNmetr}
\end{equation}
where $\>p=1+ {1\over6}\Lambda \xi\bar\xi$,
$\>q=(1-{1\over6}\Lambda\xi\bar\xi)\,\alpha+\bar\beta\,\xi+\beta\,\bar\xi$,
$\>{\rm F}=\kappa(q^2/p^2)v^2 - (q^2/ p^2)_{,u}\,v + (q/p)\,\H$,
$\>\kappa={1\over 3}\Lambda\alpha^2+2\beta\bar\beta$, and $\H=\H(\xi,\bar\xi,u)$.
The vacuum field equation is $\,p^2\,\H_{\xi\bar\xi}+{1\over 3}\Lambda\,\H=0\,$,
which has an explicit general solution given by $\H=(f_{,\xi}+\bar f_{,\bar\xi})-
{1\over3}\Lambda(\bar\xi f + \xi\bar f)/p\,$, where $f(\xi,u)$ is an
{\it arbitrary} function (otherwise (\ref{KNmetr}) is a pure radiation spacetime).
 As shown in \cite{ORR}, \cite{[A1]}, there exist the following
geometrically distinct canonical subclasses characterized by
specific choices of the parameters $\alpha$ and $\beta$:
\[ KN(\Lambda)\ \left\{\begin{array}{ll}
            \ \Lambda=0\      \left\{\begin{array}{l}
                       \kappa=0: PP\ \>\equiv KN(\Lambda=0)[\alpha=1,\beta=0]\ ,  \\
                       \kappa>0: KN\ \equiv KN(\Lambda=0)[\alpha=0,\beta=1]\ ,  \\
                       \end{array}\right. \\
                       &\\
            \ \Lambda\not=0\  \left\{\begin{array}{l}
                       \kappa>0: KN(\Lambda)I\hskip9mm\equiv KN(\Lambda)[\alpha=0,\beta=1]\ ,   \\
                       \kappa<0: KN(\Lambda^-)II\hskip4.35mm\equiv KN(\Lambda<0)[\alpha=1,\beta=0]\ ,  \\
                       \kappa=0: KN(\Lambda^-)III\hskip2.4mm\equiv
   KN(\Lambda<0)[\alpha=1,\beta=\sqrt{-\la}e^{i\omega(u)}]    \\
                        \qquad\rightarrow  KN(\Lambda^-)III_K\equiv KN(\Lambda<0)[\alpha=1,\beta=\sqrt{-\la}]
                          \ . \end{array}\right. \\
             \end{array}\right. \]
For vanishing cosmological constant $\Lambda$ there are two
subclasses. The simpler one represents the well-known  {\it pp\,}-waves denoted
here as $PP$, for which $p=1=q$, ${\rm F}=f_{,\xi}+\bar f_{,\bar\xi}\,$.
This has been investigated by many authors
(for details and references see section~21.5 in \cite{KSMH}). The  second
subclass with $\Lambda=0$ is $KN$ which is a  special class of solutions
discovered by Kundt \cite{Kundt} (see Ch.~27 in \cite{KSMH}).
Interestingly, there is an asymmetry in the case $\Lambda\ne0$:
there exist {\it three} distinct subclasses of nonexpanding waves
for $\Lambda<0$,  whereas there is only {\it one} such  subclass for $\Lambda>0$,
namely $KN(\Lambda)I$. The reason is  the condition $\kappa={1\over3}\Lambda\alpha^2+2\beta\bar\beta$
which for $\Lambda>0$ excludes the cases $\kappa<0$ and $\kappa=0$.
Note also that there exists a special subsubclass $KN(\Lambda^-)III_K$ of the
 $KN(\Lambda^-)III$ subclass such that the Debever--Penrose vector  is
the {\it Killing} vector. It was explicitly demonstrated \cite{[A1]} that this
subsubclass is identical with an interesting family of solutions found by
Siklos \cite{Sik},  and analyzed in detail in \cite{[A6]}.

The  class of expanding Robinson--Trautman solutions $RTN(\Lambda)$ has been
known for a long time \cite{RT}. In 1981, a more convenient coordinate parametrization
was found \cite{GDP}
\begin{equation}
\d s^2=2\,v^2\d\xi\d\bar\xi+2\,v\bar {\AA}\,\d\xi\d u+2\,v\AA\,\d\bar\xi
\d u+ 2\, \psi\, \d u\d v+2\,(\AA\bar {\AA}+\psi \BB)\,\d u^2\ ,
\label{RTmetr}
\end{equation}
where \ $\AA=\epsilon\xi-v f$,\ \
     $\BB=-\epsilon+{\textstyle{1\over2}}v\,(f_{,\xi}+\bar
f_{,\bar\xi})+{1\over6}\Lambda v^2\psi$,\ \
$\psi=1+\epsilon\xi\bar\xi$,\ \  and\ \  $\epsilon=-1,0,+1$.
Again, the solutions depend on an arbitrary  function $f(\xi,u)$. The parameter
$\epsilon$ is proportional to the Gaussian curvature  of the
2-surfaces $u=const$. Thus, there are nine invariant subclasses of this class
since the cases $\epsilon=+1,0,-1$ and $\Lambda>0$, $\Lambda=0$,
$\Lambda<0$ are independent.

In the subsequent paper \cite{[A2]} we presented a physical interpretation of all the
nontwisting type~{\it N\,} Einstein spaces (\ref{KNmetr}), (\ref{RTmetr}) which we based on the study of
the deviation of geodesics \cite{Pir1}. It was demonstrated that there always exists an (essentially unique)
orthonormal frame tied to any timelike observer. For the $KN(\Lambda)$ spacetimes this is given by
\begin{eqnarray}
&&e^\mu_{(0)}=\left(\dot v,\dot\xi,\dot{\bar\xi},\dot u\right)  \ ,\qquad
 e^\mu_{(1)}=-{p\over{\sqrt 2}}\left({2\,\R e\dot\xi\over q^2\dot u},1,1,0\right)
  \ ,\nonumber\\
&&e^\mu_{(2)}=-{p\over{\sqrt 2}}\left({2\,\I m\dot\xi\over q^2\dot u},i,-i,0\right)
  \ ,\qquad
e^\mu_{(3)}=-\left(\dot v-{p^2\over\dot u q^2},\dot\xi,\dot{\bar\xi},\dot u\right)
  \ ,\label{frame1}
\end{eqnarray}
and for the $RTN(\Lambda)$ solutions by
\begin{eqnarray}
&&e^\mu_{(0)}=\left(\dot v,\dot\xi,\dot{\bar\xi},\dot u\right)
  \ ,\qquad
e^\mu_{(1)}={1\over \sqrt 2\,v}\left({2v\over \psi\dot u}
  {\R e\{v\dot\xi+\AA\dot u\}},-1,-1,0\right)
  \ ,\nonumber\\
&&e^\mu_{(2)}={1\over \sqrt 2\,v}\left({2v\over \psi\dot u}
  {\I m\{v\dot\xi+\AA\dot u\}},-i,i,0\right)  \ ,\qquad
e^\mu_{(3)}=-\left(\dot v+{1\over\psi\dot u },
  \dot\xi,\dot{\bar\xi},\dot u\right)  \ .\label{frame2}
\end{eqnarray}
An invariant form of the equation of geodesic deviation  with respect to
the above interpretation frame, along any timelike geodesic
in the $KN(\Lambda)$ and $RTN(\Lambda)$ spacetimes, is  then given by
\begin{eqnarray}
\ddot Z^{(1)}&=&{\Lambda\over3}Z^{(1)}-\A_+ Z^{(1)}+\A_\times Z^{(2)} \ ,   \nonumber\\
\ddot Z^{(2)}&=&{\Lambda\over3}Z^{(2)}+\A_+Z^{(2)}+\A_\times Z^{(1)} \ , \label{E4.12}\\
\ddot Z^{(3)}&=&{\Lambda\over3}Z^{(3)}\ .\nonumber
\end{eqnarray}
The amplitudes of the  gravitational waves are given by
$\A_+=\R e\left\{\A\right\}$, $\A_\times=\I m\left\{\A\right\}$\
such that
\begin{equation}
\A=-{pq\over2}\dot u^2\> f_{,\xi\xi\xi} \label{E4.13}
\end{equation}
for the $KN(\Lambda)$  spacetimes, and
\begin{equation}
\A=- {\psi\over 2v}\dot u^2\> f_{,\xi\xi\xi} \label{E6.12}
\end{equation}
in the $RTN(\Lambda)$ spacetimes (evaluated along the geodesics). The equations
(\ref{E4.12}) express relative
accelerations of nearby free test particles in terms of their actual positions.
The frame components  of the displacement vector
$Z^{(i)}\equiv e^{(i)}_\mu Z^\mu$ determine  the
invariant distance between the test particles. Similarly,
$\ddot Z^{(i)}\equiv e^{(i)}_\mu ({D^2Z^\mu}/{d\tau^2})$
are relative accelerations. The system (\ref{E4.12})
enables us to draw the following physical conclusions:
\begin{itemize}

\item  The particles move isotropically one with respect to the other
if $\A=0$, in which case no gravitational wave is present. Indeed,
both the $KN(\Lambda)$ and $RTN(\Lambda)$ spacetimes are conformally flat
vacuum for $f_{,\xi\xi\xi}=0$, and therefore Minkowski ($\Lambda=0$),
de Sitter ($\Lambda>0$), and anti-de Sitter ($\Lambda<0$).
These maximally symmetric, homogeneous, and isotropic spacetimes
are thus natural backgrounds for other ``non-trivial''
$KN(\Lambda)$ and $RTN(\Lambda)$  radiative solutions.

\item If $f_{,\xi\xi\xi}\not=0$ then the amplitudes $\A$ do not
vanish. The particles are influenced by the wave, in a similar way as
 they are affected by a standard gravitational wave
in Minkowski background. For $\Lambda\not=0$,
the influence of the  wave is added to the (anti-)de Sitter isotropic
expansion/contraction. Therefore,  the $KN(\Lambda)$ and $RTN(\Lambda)$ metrics
can be interpreted as exact gravitational waves which propagate in spaces of
constant curvature.

\item The wave has a transverse character as only motions in the
plane ($\e_{(1)}, \e_{(2)}$) perpendicular to $\e_{(3)}$ are affected. The
propagation direction is $\e_{(3)}$ and coincides with the projection of
the Debever--Penrose vector  on the  hypersurface orthogonal to the observer's
four-velocity.

\item There are two polarization modes of the wave,  ``+''  and ``$\times$'',
with the corresponding amplitudes  $\A_+$ and $\A_\times$.
Under rotations in the transverse plane they transform according to
$\A' _+= \cos 2\vartheta \A_+ -\sin 2\vartheta \A_\times$,
$\A' _\times= \sin 2\vartheta \A_+ +\cos 2\vartheta \A_\times$,
so that the helicity of the wave is 2, as with linearized
waves on a Minkowski background.

\item Curvature singularities (which might be considered as
``sources'' of the gravitational waves) occur where the amplitudes $\A$ diverge.
Note that the singularities of the $RTN(\Lambda)$ spacetimes can  be
characterized by the non-vanishing invariant constructed recently
by Bi\v c\'ak and Pravda \cite{BPr} from the second derivatives of the Riemann tensor.

\item  Special classes of explicit geodesics were also discussed in
\cite{[A2]}. For a {\it positive} cosmological constant, the amplitudes $\A$
decay exponentially fast, i.e. the {\it waves are damped}.
The spacetimes locally approach the de Sitter universe, which is
an explicit demonstration of the  cosmic no-hair  conjecture \cite{Maeda} under
the presence of waves in exact model spacetimes. With Ji\v r\'\i\
Bi\v c\'ak we also discussed the validity of  the cosmic no-hair conjecture
in the Robinson--Trautman radiative spacetimes of Petrov type
{\it II\ }\ \cite{[A7]}.

\end{itemize}

\subsection{Motivation: impulses as limits of exact sandwich waves}

We have already indicated at the beginning of previous section that probably the most natural way
how to introduce impulsive waves in spaces of constant curvature is to consider a distributional
limit of sandwich waves of the above $KN(\Lambda)$ and $RTN(\Lambda)$ spacetimes.
Indeed, all these solutions depend, in suitable coordinates (\ref{KNmetr}), (\ref{RTmetr}),
on an arbitrary function $f(\xi,u)$ where the coordinate $u$ is the ``retarded time".
Thus, an arbitrary wave-profile  can be prescribed within the above families of exact radiative
solutions. In particular, we can  consider sandwich waves  for which the wave-profiles
 $d_\varepsilon(u)$ of the function $f(\xi,u)\equiv f(\xi)\,d_\varepsilon(u)$
are nonvanishing only on a {\it finite interval}  of $u$ (of the length $\varepsilon$ around $u=0$).
In the context of the simplest class of {\it pp\,}-waves this has been introduced already in
the late 50's by Bondi, Pirani  and Robinson \cite{BPR}.

Now, the distributional limit $d_\varepsilon(u)\to \delta(u)$, where $\delta$ is the Dirac
delta distribution, results (at least formally) in a solution which clearly
represents an impulsive gravitational wave localized on a single wave-front $u=0$.
Such a procedure, in which a sequence of sandwich {\it pp\,}-waves with profiles
$d_\varepsilon(u)$ becomes infinitesimal in duration but unbounded in the amplitude
as $\varepsilon\to0$, was considered  in  \cite{Pen68}, \cite{Rindler} and later
elsewhere \cite{[A3]}. In this simplest case given by the metric (\ref{KNmetr}) with
$p=1=q$, ${\rm F}=\H(\xi,\bar\xi)\,d_\varepsilon(u)$, one directly obtains the metric
 \begin{equation}
\d s^2=2\,\d\xi\,\d\bar\xi - 2\,\d u\,\d v+\H\,\delta(u)\, \d u^2
\ ,\label{imp-pp}
\end{equation}
which is  the well-know Brinkmann form of impulsive plane  wave in  Minkowski background
\cite{Pen68}, \cite{Pen72}. Analogously, more general nonexpanding impulsive waves of the Kundt class
$KN(\Lambda)$ for $\Lambda\ne0$, and expanding Robinson--Trautman
impulsive waves of the $RTN(\Lambda)$ class can be constructed (see  section 3.3
below for more details).

Mathematical problems and questions arise however with the above procedure. What is the meaning
of the metric with distributional terms? What is the manifold  structure?
Usually, the metric coefficients are considered to be $C^2$ continuous functions.
And even worse: in the limit of the $RTN(\Lambda)$ sandwich-wave spacetimes one obtains
a metric which contains a term proportional to $\delta^2$. The
product of the Dirac distributions is a mathematically ill-defined object within
the Schwartz linear distribution theory. For this, it may be necessary to invoke the
more advanced  Colombeau nonlinear theory of generalized functions \cite{Colomb}.

In view of these (and other) open questions, it is of primary importance to investigate
the  construction of impulsive-wave spacetimes by {\it various} methods and
approaches. These may provide a deeper understanding of the impulsive solutions based
on mathematically sound definitions, discover mutual relations, investigate  specific
properties, and elucidate their physical interpretation.

\section{Main theme: methods of construction of impulsive waves}

The history of studies of impulsive waves can be divided
(roughly speaking) into three  periods. During the first epoch, which
culminated in classic works by Lichnerowicz and others \cite{Lich},
 principal mathematical properties of possible shock or
impulsive waves in general relativity were found and investigated. It was
demonstrated  that these  must necessarily be localized on
null hypersurfaces, across which the derivatives of the metric
(the second or the first) are discontinuous.
Nevertheless, explicit constructions of  impulsive solutions and further
investigation of their properties had not been performed
until the fundamental contributions by Penrose \cite{Pen72} and Aichelburg
and Sexl \cite{AS} appeared at the beginning of 1970's.
During this ``Golden Era'' most of the different methods of construction
occured almost simultaneously. These methods, based on various
approaches, are summarized in the following table:

\bigskip
\begin{tabular}{|l|c|}
\hline
 {\it \ \ method of construction} &
 {\it basic characteristics} \\
\hline
\hline
  ``cut and paste'' & general but only ``formal''\\
\hline
  continuous coordinates   & general and  explicit\\
\hline
 limits of sandwich waves   & general,  explicit (problems for expanding impulses)\\
\hline
  embedding  & special (nonexpanding $\Lambda\ne0$ impulses only), explicit\\
\hline
  boosts/infinite acceleration & special (but physically important) and explicit\\
\hline
\end{tabular}
\bigskip

\noindent

The last era, which can be called the ``Renaissance'', started
at the beginning of 1990's with a fundamental paper by Hotta and Tanaka
\cite{HotTan93}. By  boosting the Schwarzschild--(anti-)de~Sitter black hole
to the speed of light they  constructed (after previous ``no-go'' attempts
by several authors) a nonexpanding impulse in the de~Sitter and anti-de~Sitter
universes. Simultaneously, an interesting expanding impulsive-wave solution
generated by a ``snapping'' cosmic string in Minkowski space was discovered
and discussed by Gleiser and Pullin \cite{GlePul89}, Bi\v{c}\'ak \cite {Bi90},
Nutku and Penrose \cite{NutPen92}, and Hogan \cite{Hogan93}, \cite{Hogan94}.

As a result of these and subsequent systematic investigations an almost complete
picture has now emerged. This comprises both nonexpanding and expanding impulses in
constant-curvature spaces with an arbitrary value of the cosmological constant.
We will describe  the classic and also more recent results here. We shall present
these in the context of a unifying description of the main methods by which  complete
families of impulsive gravitational waves can be constructed. We will also
mention the corresponding references and  historical remarks in some more detail.

\subsection{The ``cut and paste'' method}

Let us start with an elegant  geometrical method for constructing general
nonexpanding (plane) and expanding (spherical) impulsive gravitational waves in
Minkowski background. This was presented by Penrose in now classic works \cite{Pen68},
\cite{Pen72}. His ``cut and paste'' approach, which is in some respects similar
to that of Israel \cite{Israel}, is based on  cutting the spacetime manifold ${\cal M}$
along a suitable null hypersurface and then re-attaching the two pieces with a
specific ``warp''. The first case leads to impulsive {\it pp\,}-waves, the
second case gives expanding impulsive waves. Recently,
this approach has also been extended to constant-curvature backgrounds with a nonzero
cosmological constant $\Lambda$ \cite{Hogan92}-\cite{[B10]}.

\subsubsection{ Nonexpanding impulsive waves}

In this case, the Penrose method is based on the removal of the null hypersurface
$\U=0$ from spacetime in the metric form  (\ref{conf})
 \begin{equation}
\d s_0^2= {2\,\d\eta\,\d\bar\eta -2\,\d\U\,\d\V
\over [\,1+{1\over6}\Lambda(\eta\bar\eta-\U\V)\,]^2}\ ,
 \label{conf*}
 \end{equation}
and re-attaching the halves ${\cal M}_-(\U<0)$ and  ${\cal M}_+(\U>0)$
by making the identification with a ``warp'' in the coordinate $\V$ such that
\begin{equation}
\Big[\,\eta,\,\bar\eta,\,\V,\,\U=0_-\,\Big]_{_{{\cal M}_-}}\equiv
\Big[\,\eta,\,\bar\eta,\,\V-H(\eta,\bar\eta),\,\U=0_+\,\Big]_{_{{\cal M}_+}}\ , \label{juncnon}
\end{equation}
where $H(\eta,\bar\eta)$ is an arbitrary function of $\eta$ and $\bar\eta$ alone.
It was shown in \cite{Pen72} that the condition (\ref{juncnon}) automatically guarantees that
the Einstein field equations are satisfied everywhere including on $\>\U=0$.
However, impulsive components are introduced into the curvature tensor
proportional to $\delta(\U)$. These represent gravitational (plus
possibly null-matter) impulsive waves.

In  \cite{Pen72} Penrose considered  only the Minkowski background space
(\ref{conf}) with $\Lambda=0$, in which case the impulsive surface
$\>\U=0$ is a plane. Impulsive {\it pp\,}-waves are thus
obtained. However, the above  method can easily be
generalized to $\Lambda\ne0$ cases. It was shown in \cite{[B7]}
that exactly the same junction conditions (\ref{juncnon}) applied to a
general background spacetime (\ref{conf*}) introduce impulsive waves
also in the de Sitter ($\Lambda>0$) or anti-de~Sitter ($\Lambda<0$)
universes. However, the geometries of these impulses are different as
the null hypersurface $\>\U=0$, along which the spacetime is cut and
pasted, is described by the 2-metric
$\d\sigma^2= 2\,(\,1+{1\over6}\,\Lambda\,\eta\,\bar\eta\,)^{-2}\,\d\eta\,\d\bar\eta $.
This is  obviously a 2-dimensional space of constant Gaussian curvature $K={1\over3}\Lambda$.
When $\Lambda=0$ the impulsive wave surface is a {\it plane}. For $\Lambda>0$ it is
a {\it sphere}, while for $\Lambda<0$ it is a {\it hyperboloid}. For cases
$\Lambda\ne0$, the geometry of these  impulsive spherical and hyperboloidal
waves was described in detail in~\cite{[B5]} using various convenient
coordinate representations. It was also demonstrated that the wave
surfaces are indeed nonexpanding.

Let us remark that  the Penrose ``cut and paste'' approach to
construction of nonexpanding impulsive waves is closely related to
the Dray and 't~Hooft method of ``shift function'', which is prescribed on
the horizon of a suitable background space \cite{DaT}.

\subsubsection{ Expanding impulsive waves}

In  \cite{Pen72} Penrose demonstrated that  his ``cut and paste'' method
can be also used to construct expanding spherical impulsive (purely) gravitational
waves. However, the junction conditions are somewhat more complicated than in the
nonexpanding case. Instead of starting with the familiar
form of  Minkowski or (anti-)de~Sitter spacetime (\ref{conf}), it is  necessary
to first perform the transformation
\begin{equation}
  \U= {Z\bar Z\over p}\,V-U\ , \qquad
  \V= {V\over p}-\epsilon U\ , \qquad
  \eta= {Z\over p}\,V\ ,
  \label{inv}
\end{equation}
  where
\begin{equation}
  p=1+\epsilon Z\bar Z\ , \qquad \epsilon=-1,0,+1\ .
  \label{pandep}
\end{equation}
With this, the constant curvature spacetimes take the following form
  \begin{equation}
\d s_0^2 = {{ \displaystyle
  2\,(V/p)^2\,\d Z\,\d\bar Z +2\,\d U\,\d V -2\epsilon\,\d U^2}\over
\big[\,1+{1\over6}\Lambda U(V-\epsilon U)\,\big]^2}\ .
  \label{U<0}
  \end{equation}
In these coordinates, the hypersurface  $U=0$ is a {\it null cone} (a sphere expanding with the
speed of light) as $\>\eta\bar\eta-\U\V\equiv U(V-\epsilon U)=0$.
The spacetime can now be divided into two halves
${\cal M}_-(U<0)$ inside the null cone, and  ${\cal M}_+(U>0)$ outside this.
Now, the Penrose junction conditions prescribe the identification
\begin{equation}
\Big[\,Z,\,\bar Z,\,V,\,U=0_-\,\Big]_{_{{\cal M}_-}}\equiv
\Big[\,h(Z),\,\bar h({\bar Z}),
\,{(1+\,\epsilon h\bar h\,)\,V\,\,\over(1+\epsilon Z\bar Z)|h'|},\,U=0_+\,\Big]_{_{{\cal M}_+}}
\  \label{juncexp}
\end{equation}
of the  points from the two re-attached parts across the impulsive  sphere $U=0$.
In (\ref{juncexp}) an arbitrary function $h(Z)$ introduces a specific ``warp''
corresponding to one of all possible impulsive {\it vacuum} spacetimes of this type.

In \cite{Pen72} Penrose described  the above ``cut and paste'' construction
 for the case of expanding spherical gravitational waves in
a Minkowski background. He assumed  the simplest case $\Lambda=0$,
$\epsilon=0$. Generalization to impulsive spherical waves in the
de Sitter and anti-de~Sitter universe (with $\epsilon=0$) was
later found by Hogan \cite{Hogan92}. In \cite{Hogan94} he also
introduced impulses in Minkowski spacetime with $\epsilon=+1$.
The completely general form (\ref{U<0}), (\ref{juncexp})
of the Penrose junction conditions has
been recently presented in our contributions \cite{[B8]}, \cite{[B10]}.

Note finally that a similar (yet somewhat different) ``cut and paste'' approach
was considered by Gleiser and Pullin \cite{GlePul89} for constructing a
specific solution which represents a spherical impulsive gravitational wave
generated by a ``snapping'' cosmic string in Minkowski space. This
particular solution will be discussed below in subsection 3.5.2.

\subsection{Continuous coordinates}

The Penrose ``cut and paste'' approach described in the previous
section 3.1 is a general method. By prescribing the junction conditions
(\ref{juncnon}) or (\ref{juncexp}) in  (\ref{conf*}) or (\ref{U<0}), respectively,
all nonexpanding and expanding impulsive
gravitational waves in Minkowski, de~Sitter, and anti-de~Sitter
universe can be constructed. Nevertheless, these formal identifications of points
on both sides of the impulsive hypersurface do not immediately provide
explicit metric forms of the complete spacetimes. It is thus of
primary interest to find a suitable coordinate system for the
above solutions in which the metric is explicitly continuous
everywhere, including on the impulse. In the following we describe
the construction of these privileged continuous coordinates.

\subsubsection{ Nonexpanding impulsive waves}

We start again with the metric (\ref{conf}) which represents spaces of constant
curvature. Let us  perform the transformation $\U=U$, $\V=V+H+UH_{,Z}H_{,\bar Z}$,
and $\eta=Z+UH_{,\bar Z}$, where $H(Z,\bar Z)$ is an arbitrary real function.
This yields the metric
 \begin{equation}
\d s_0^2= {2\,|\d Z+U(H_{,Z\bar Z}\d Z+H_{,\bar Z\bar Z}\d\bar Z)|^2-2\,\d U\d V
\over [\,1+{1\over6}\Lambda(Z\bar Z-UV-UG)\,]^2}\ ,
 \label{deS2}
 \end{equation}
 where $G=H-ZH_{,Z}-\bar ZH_{,\bar Z}$.
We now consider (\ref{deS2}) for $U>0$ and combine this with the line element
(\ref{conf}) in which we take $\U=U$, $\V=V$, $\eta=Z$ for $U<0$. The resulting
metric can be written as
 \begin{equation}
\d s^2= {2\,|\d Z+U\Theta(U)(H_{,Z\bar Z}\d Z+H_{,\bar Z\bar Z}\d\bar Z)|^2-2\,\d U\d V
\over [\,1+{1\over6}\Lambda(Z\bar Z-UV-U\Theta(U)G)\,]^2}\ ,
 \label{conti}
 \end{equation}
 where $\Theta(U)$ is the Heaviside step function. This is continuous across the
null hypersurface $U=0$ but the discontinuity in the derivatives of the metric
introduces  impulsive components in the Weyl and curvature tensors proportional
to the Dirac  distribution \cite{[B7]},
$\>\Psi_4 = (1+{1\over6}\Lambda Z\bar Z)^2 H_{,ZZ} \,\delta(U)\,$,
$\> \Phi_{22} = [\, (1+{1\over6}\Lambda Z\bar  Z )^3
 ((1+{1\over6}\Lambda Z\bar Z)^{-1}H)_{,Z\bar Z}
+{1\over3}\Lambda \,H\, ]\, \delta(U)\,$, in a natural tetrad.
The metric (\ref{conti}) thus explicitly describes  impulsive
waves in de~Sitter, anti-de~Sitter or Minkowski backgrounds. For $\Lambda=0$,
the line element (\ref{conti}) reduces to the well-known Rosen form of impulsive
{\it pp\,}-waves \cite{Pen72}, \cite{[B2]}, \cite{Steinb}. Note also
that the continuous coordinate system for the particular
Aichelburg--Sexl solution \cite{AS} (see (\ref{AiSe}) further in the text)
was found by D'Eath \cite{DE78} and used
for analytic investigation of ultrarelativistic black-hole encounters. This
was used for studies of high-energy scattering in quantum
 gravity on the Planck scale \cite{Ver}.

The complete transformation which relates (\ref{conf}) and (\ref{conti})
can be written as
 \begin{eqnarray}
\U&=&U\ , \nonumber\\
\V&=&V+H\,\Theta(U)+U\,\Theta(U)\,H_{,Z}H_{,\bar Z}\ ,  \label{trans}\\
\eta&=&Z+U\,\Theta(U)\,H_{,\bar Z}\ , \nonumber
 \end{eqnarray}
which is discontinuous in the coordinate $\V$ on $\U=0$. From (\ref{trans}), using the fact that the
coordinates $U,V, Z$ are continuous, we obtain  exactly the Penrose junction condition (\ref{juncnon})
for reattaching the two halves of the spacetime ${\cal M}_-$ and ${\cal M}_+$ with a warp.
Thus, the above procedure is indeed an {\it explicit} Penrose's ``cut and paste''
construction of all nonexpanding impulsive gravitational waves.

Of course, the discontinuity in the complete transformation (\ref{trans}),
which (formally) relates the continuous to the distributional form
of  impulsive solutions, causes some  mathematical problems. However,
it has been recently shown in \cite{Steinb}, using the previous results
\cite{Steina}, that (\ref{trans}) is in fact  a {\it rigorous} example of a
generalized coordinate transformation in the sense of
Colombeau's generalized functions (at least in the case of impulsive {\it pp\,}-waves).
Moreover, it is possible to
interpret this change of coordinates as the distributional limit
of a family of smooth transformations which is obtained by a
general regularization procedure, i.e. by considering the
impulse as a limiting case of sandwich waves with an {\it arbitrarily}
regularized wave profile. These results put the formal (``physical'')
equivalence of  both continuous and distributional forms of impulsive spacetimes
on a solid ground.

\subsubsection{ Expanding impulsive waves}

In this case, we perform a more involved transformation of the metric (\ref{conf})
given by
 \begin{eqnarray}
 \V&=& AV-DU\ ,   \nonumber\\
 \U&=& BV-EU\ ,   \label{transe3} \\
 \eta&=& CV-FU\ ,  \nonumber
 \end{eqnarray}
where
 \begin{eqnarray}
&&A= \frac{1}{p|h'|}\ ,\qquad
B= \frac{|h|^2}{p|h'|}\ ,\qquad
C= \frac{h}{ p|h'|}\ ,   \nonumber\\
&&D= \frac{1}{|h'|}\left\{
\frac{p}{4} \left|\frac{h''}{h'}\right|^2+\epsilon
\left[1+\frac{Z}{2}\frac{h''}{h'}+\frac{\bar Z}{2}\frac{\bar h''}{\bar h'}
\right]\right\}\ , \label{transe4}\\
&&E= \frac{|h|^2}{|h'|}
\bigg\{ \frac{p}{4}\left|\frac{h''}{h'}-2\frac{h'}{h}\right|^2
+\epsilon\left[ 1+\frac{Z}{2}
\left(\frac{h''}{h'}-2\frac{h'}{h}\right)+\frac{\bar Z}{2}
\left(\frac{\bar h''}{\bar h'}-2\frac{\bar h'}{\bar h}\right)
\right]\bigg\}\ ,\nonumber\\
&&F= \frac{h}{|h'|}\bigg\{
\frac{p}{4}\left(\frac{h''}{h'}-2\frac{h'}{h}\right)
\frac{\bar h''}{\bar h'}
+\epsilon\left[1+
 \frac{Z}{2}\left(\frac{h''}{h'}-2\frac{h'}{h}\right)
+\frac{\bar Z}{2}\frac{\bar h''}{\bar h'}\right]\bigg\}\ .\nonumber
 \end{eqnarray}
Here  $h=h(Z)$ is an arbitrary function and the derivative with respect to
its argument $Z$ is denoted by a prime. With this, the  metric
(\ref{conf}) of a constant curvature space becomes
 \begin{equation}
 \d s_0^2 = {2\left| (V/p)\,\d Z+U\,p\,\bar H\,\d\bar Z \right|^2
  +2\,\d U\,\d V -2\epsilon\,\d U^2\over
\big[\,1+{1\over6}\Lambda U(V-\epsilon U)\,\big]^2}\ ,
 \label{U>0}
 \end{equation}
 where
 \begin{equation}
 H(Z)={1\over2} \left[\,{h'''\over h'}-{3\over2}\left({h''\over  h'}\right)^{\!2}
 \,\right] .  \label{Schwarz}
 \end{equation}
Similarly as in the nonexpanding case, we may now combine the line element (\ref{U>0})
for $U>0$ and attach this to the metric (\ref{U<0}) for $U<0$ (which was obtained from
(\ref{conf}) by the transformation (\ref{inv})). The resulting complete line element can
thus be written in the form
 \begin{equation}
\d s^2 =  {2\left| (V/p)\,\d Z+U\Theta(U)\,p\,\bar H\,\d\bar Z \right|^2
+2\,\d U\,\d V -2\epsilon\,\d U^2\over
\big[\,1+{1\over6}\Lambda U(V-\epsilon U)\,\big]^2}\ .
 \label{en0}
 \end{equation}
 This  metric which was presented for a Minkowski background in
\cite{NutPen92}-\cite{Hogan94}, with a cosmological constant in
\cite{Hogan92}, and in the most general form in \cite{[B8]}, \cite{[B10]}
 is explicitly continuous everywhere, including across the null
hypersurface $U=0$. Again, the discontinuity in the derivatives of the
metric yields impulsive components in the Weyl and curvature
tensors,
$\Psi_4=(p^2H/ V)\,\delta(U)$,
$\Phi_{22}=(p^4H\bar H/ V^2)\,U\,\delta(U)$,
using a naturally adapted tetrad.
The spacetime is clearly conformally flat everywhere except on the impulsive
wave surface $U=0$. It is a vacuum solution everywhere except on the wave
surface at $V=0$, and at possible singularities of the function $p^2H$.

By comparing the transformations (\ref{inv}) at $U=0_-$ with
(\ref{transe3}), (\ref{transe4}) at $U=0_+$, we  obtain exactly the Penrose
junction conditions (\ref{juncexp}).
The above procedure is thus an {\it explicit} ``cut and paste'' construction
of expanding spherical pure gravitational waves in spaces of constant curvature.

Note also that  new continuous coordinates which generalize (\ref{en0}) for the
case $\Lambda=0$ were found recently \cite{NutAli} as an extension of previous results for
spherical shock waves \cite{Nutku}. These contain an
additional parameter which is related to  acceleration of the
coordinate system.

\subsection{Limits of sandwich waves}

We have already mentioned in the introductory section~2 that impulsive waves can be understood as
distributional limits of  appropriate sequence of sandwich waves in a suitable family of exact
radiative spacetimes. In fact,  this seems to be the most intuitive way of construction of
impulsive waves, although mathematical difficulties occur with this approach in the case of
expanding impulses of the Robinson--Trautman type.

\subsubsection{ Nonexpanding impulsive waves}

It has been demonstrated in \cite{[B1]} that {\it all} nonexpanding
impulses in Minkowski, de~Sitter or anti-de~Sitter universes can
simply be constructed from the Kundt class $KN(\Lambda)$ of exact type {\it N}
solutions (\ref{KNmetr}) by considering the distributional form
$\H(\xi,\bar\xi)\,\delta(u)$ of the structural function $\H$. These metrics
can thus be written as
\begin{equation}
\d s^2=2\,{1\over p^2}\,\d\xi\d\bar\xi - 2\ {q^2\over p^2}\,\d
u\d v+ \left[\kappa{q^2\over p^2}\,v^2 - {(q^2)_{,u}\over p^2}\,v + {q\over p}
  \,\H(\xi,\bar\xi)\,\delta(u)\right]\, \d u^2\ ,\label{KNimp}
\end{equation}
where $\>p=1+ {1\over6}\Lambda \xi\bar\xi$,
$\>q=(1-{1\over6}\Lambda\xi\bar\xi)\,\alpha+\bar\beta\,\xi+\beta\,\bar\xi$,
and $\>\kappa={1\over 3}\Lambda\alpha^2+2\beta\bar\beta$.
As indicated in section 2.2, for a general wave-profile of the function $\H$ there
exist various distinct canonical subclasses of (\ref{KNmetr}) characterized by
specific choices of the parameters $\alpha$ and $\beta$. However, {\it impulsive}
limits of these  subclasses become (locally) {\it equivalent}. Indeed, in the
case of vanishing cosmological constant $\Lambda$ there are two
subclasses, namely the  {\it pp\,}-waves $PP$ and the Kundt subclass  $KN$.
However,  the transformation
\begin{equation}
   \U    = (\xi+\bar\xi)(1+uv)u\ , \qquad
   \V+1  = (\xi+\bar\xi)v      \ , \qquad
   \eta  = \xi +(\xi+\bar\xi)uv \ , \label{E3.1}
\end{equation}
converts the impulsive $KN$ metric (\ref{KNimp}) with $\alpha=0$, $\beta=1$ to
the impulsive $PP$ metric with $\alpha=1$, $\beta=0$.
The only non-trivial impulsive gravitational waves
of the form (\ref{KNimp}) in Minkowski space are thus impulsive
{\it pp\,}-waves (\ref{imp-pp}).

Similar results hold also for the $\Lambda\ne0$ case. Generically, there are
three distinct subclasses of nonexpanding waves $KN(\Lambda)I$,
$KN(\Lambda^-)II$, and $KN(\Lambda^-)III$ (see section 2.2).
It was shown in \cite{[B1]} that although these canonical
subclasses  are different for smooth profiles, they are  equivalent for
impulsive profiles. For example, the metric of the $KN(\Lambda^-)I$
subclass of (\ref{KNimp}) given by $\alpha=0$, $\beta=1$,  $\Lambda<0$
 can be expressed, using the transformation
$ \xi = \sqrt{2} a\, e^{i\phi_1}\tanh(R_1/2)$,
$  v = {1\over\sqrt{2}}\,a/t_1$,
$   u = {1\over \sqrt{2}}\, (\rho_1-t_1)/a$, as
\begin{eqnarray}
&&\d s^2={a^2\over t_1^2}\sinh^2 R_1\cos^2 \phi_1\,
 (\d\rho_1^2-\d t_1^2)+a^2(\d R_1^2+\sinh^2 R_1\d \phi_1^2) \nonumber\\
 &&\qquad\qquad\qquad +\sinh R_1\cos\phi_1\, \H(R_1,\phi_1)
   \delta(t_1-\rho_1)(\d t_1-\d\rho_1)^2 \ .
  \label{E5.9}
 \end{eqnarray}
Similarly, the transformation
$\> \xi = \sqrt{2} a\, e^{i\phi_2}\tanh(R_2/2)\,$,
$\> v = -a^2/\rho_2\,$,
$\> u = t_2 - \rho_2\>$
brings the impulsive subclass  $KN(\Lambda^-)II$ of the metric (\ref{KNimp}) with $\alpha=1$,
$\beta=0$, $\Lambda<0$ to the form
\begin{eqnarray}
&&\d s^2={a^2\over \rho_2^2}\cosh^2 R_2\,
 (\d\rho_2^2-\d t_2^2)+a^2(\d R_2^2+\sinh^2 R_2\d \phi_2^2) \nonumber\\
 &&\qquad\qquad\qquad +\cosh R_2\, \H(R_2,\phi_2)
   \delta(t_2-\rho_2)(\d t_2-\d\rho_2)^2 \ .
  \label{E5.12}
 \end{eqnarray}
However, the  two metrics (\ref{E5.9}) and (\ref{E5.12}) are
equivalent under the transformation
 \begin{eqnarray}
&& t_1^2  = t_2^2-\rho_2^2(1-\tanh^2 R_2\cos^2\phi_2)\ , \qquad
 \rho_1   = \rho_2\tanh R_2\cos\phi_2 \ ,  \nonumber\\
&& \cosh R_1 = {t_2\over \rho_2}\cosh R_2 \ , \qquad
 \sin\Phi_1  = -{\sinh R_2\sin\phi_2\over    \sqrt{t_2^2\rho_2^{-2}\cosh^2 R_2-1}} \ .
 \label{E5.20}
 \end{eqnarray}

In this way, by considering the above distributional limit (\ref{KNimp}) of the
$KN(\Lambda)$ class  (\ref{KNmetr}) we obtain an explicit form of   solutions
which represent nonexpanding impulses. The metric (\ref{KNimp}) contains a single
Dirac delta in the $\d u^2$ term, and obviously generalizes the well-known
Brinkmann form of impulsive {\it pp\,}-waves
(\ref{imp-pp}) to which it reduces when $p=1=q$.

Interestingly, there  exists yet another simple coordinate form of the complete
family of nonexpanding impulses of the Kundt class $KN(\Lambda)$
which also contains a term with the Dirac delta. This is obtained from
the continuous form of the impulsive-wave metric (\ref{conti}) by
applying the transformation (\ref{trans}). If we take into account the
terms which arise from the derivatives of $\Theta(\U)$, this transformation
relates   (\ref{conti}) to
 \begin{equation}
\d s^2= {2\,\d\eta\,\d\bar\eta-2\,\d \U\,\d \V
+2H(\eta,\bar\eta)\,\delta(\U)\,\d \U^2
\over [\,1+{1\over6}\Lambda(\eta\bar\eta-\U\V)\,]^2}\ ,
 \label{confppimp}
 \end{equation}
 which explicitly includes the impulse  located on the wavefront $\U=0$.
 This is a gravitational wave or an impulse of
null matter  in any  spacetime of a constant curvature (\ref{conf}),
depending on the specific form of the function $\H$ (see  \cite{[B7]}, \cite{HorItz99}
for more details). Again, in Minkowski background
this is just the Brinkmann form of a general impulsive
{\it pp\,}-wave (\ref{imp-pp}). In fact, the metric
(\ref{confppimp}) is conformal to (\ref{imp-pp}). (Recall in this context that
Siklos  \cite{Sik} proved that Einstein spaces conformal to {\it pp\,}-waves
only occur when $\Lambda<0$. The impulsive case is clearly a counter-example to
this result.)

The explicit transformation between (\ref{KNimp}) and
(\ref{confppimp}) for the $KN(\Lambda)I$ subclass is
\begin{eqnarray}
&&   \U    = {(\xi+\bar\xi)(1+uv)u\over
  1-{1\over6}\Lambda(\xi+\bar\xi)(1+uv)u}\ , \nonumber\\
&&   \V  = {(\xi+\bar\xi)v+{1\over6}\Lambda\xi\bar\xi\over
  1-{1\over6}\Lambda(\xi+\bar\xi)(1+uv)u} -1     \ , \label{tr}\\
&&   \eta  = {\xi +(\xi+\bar\xi)uv\over
  1-{1\over6}\Lambda(\xi+\bar\xi)(1+uv)u} \ , \nonumber
\end{eqnarray}
which reduces to (\ref{E3.1}) in the case $\Lambda=0$. Similar
transformations hold also for the subclasses $KN(\Lambda^-)II$ and
$KN(\Lambda^-)III$. Therefore, the full family of impulsive limits (\ref{KNimp})
of  nonexpanding sandwich waves of the Kundt class $KN(\Lambda)$
is indeed equivalent to the distributional form of the solutions
 (\ref{confppimp}), and consequently to the continuous metric (\ref{conti})
 obtained by the explicit ``cut and paste'' method.

\subsubsection{ Expanding impulsive waves}

As has been argued in \cite{[B8]},  the family of solutions
for expanding impulsive spherical gravitational waves can be considered to
be an impulsive limit of the class $RTN(\Lambda)$ of vacuum Robinson--Trautman
type~$N$ solutions with a cosmological constant. It is
convenient to consider these solutions in the Garc\'{\i}a--Pleba\'nski
\cite{GDP} coordinates (\ref{RTmetr}). Introducing a new coordinate $w$ by
$w=\psi\, v$, and assuming the impulsive limit $f\equiv f(\xi)\delta(u)$, we can
express the above family of expanding impulses as
  \begin{eqnarray}
  \d s^2 &=& 2{w^2\over\psi^2}\,
\Big|\,\d\xi-f\,\delta(u)\,\d u\Big|^2 +2\,\d u\,\d w
+\Big({\textstyle{1\over3}}\Lambda w^2-2\epsilon\Big)\d u^2
\nonumber\\
&&\qquad\qquad  +w\,\Big[ (f_{,\xi}+\bar f_{,\bar\xi})
-{2\epsilon\over\psi}(f\bar\xi +\bar f\xi)
\Big] \delta(u) \d u^2\ ,
  \label{RTNLe}
  \end{eqnarray}
  where $f(\xi)$ is an arbitrary function. Similarly as in the nonexpanding case
(\ref{KNimp}),  the spherical impulse in the form (\ref{RTNLe}) is explicitly
located on the null  surface  $u=0$. For $u\ne0$ the
spacetime again reduces to the background  of constant curvature.
However, a difficult mathematical problem occurs in this case: the metric
(\ref{RTNLe}) contains the {\it product} of the Dirac distributions in
the first term. This is  not a well-defined concept in the linear theory
of distributions, so that the metric (\ref{RTNLe}) is only formal.
Nevertheless, it demonstrates that  impulsive limits of Robinson--Trautman
type $N$ vacuum spacetimes are equivalent to expanding impulsive
gravitational waves (\ref{en0}). Indeed,  the transformation
 \begin{equation}
 w = {\tilde w\over1+{1\over6}\Lambda\tilde u(\tilde w-\epsilon \tilde u)}\ , \qquad
 u = \int{\d \tilde u\over1-\epsilon{1\over6}\Lambda\tilde u^2}\ ,
 \label{trans2}
 \end{equation}
converts  (\ref{RTNLe}) to
   \begin{eqnarray}
  \d s^2 &=& {1\over[\,1+{1\over6}\Lambda\tilde u(\tilde w-\epsilon \tilde u)\,]^2}
  \Bigg[
  2{\tilde w^2\over\psi^2}\,
\Big|\,\d\xi-f\,\delta(\tilde u)\,\d \tilde u\Big|^2 +2\,\d \tilde u\,\d \tilde w
-2\epsilon\,\d \tilde u^2 \nonumber\\
&&\hskip40mm  +\tilde w\,\Big[ (f_{,\xi}+\bar f_{,\bar\xi})
-{2\epsilon\over\psi}(f\bar\xi +\bar f\xi) \Big] \delta(\tilde u) \d \tilde u^2\Bigg]\ .
  \label{modif}
  \end{eqnarray}
Dropping the tildes  and performing the following discontinuous transformation
\begin{eqnarray}
 u&=& U+\Theta(U)  \,[ U_{inv}(U,V,Z,\bar Z)-U\,]\ , \nonumber\\
 w&=& V+\Theta(U)  \,[ V_{inv}(U,V,Z,\bar Z)-V\,]\ , \label{transe}\\
 \xi&=& Z+\Theta(U)\,[Z_{inv}(U,V,Z,\bar Z)-Z \,]\ , \nonumber
 \end{eqnarray}
where the functions $U_{inv}$, $V_{inv}$, $Z_{inv}$ are obtained by the composition
of the inverse transformation to (\ref{inv}) with (\ref{transe3}), we put (\ref{modif})
into the metric form of all expanding impulsive spherical waves expressed in the continuous
coordinate system (\ref{en0}) (see \cite{[B8]} for more details). This relation is obvious
for all $U\not=0$ if we use the identity  $u(w-\epsilon u)= U(V-\epsilon U)$. Keeping
the distributional terms arising from $\Theta$ and its derivative, we formally obtain also
the impulsive terms proportional to $\delta$, with the identification
$f\equiv Z_{inv}-Z$ (evaluated on $U=0$).

Of course, much work is required before the metric (\ref{RTNLe}) and the
discontinuous transformation (\ref{transe}) can be fully accepted
in a mathematically rigorous way. A natural tool appears to be the
Colombeau theory of nonlinear generalized functions \cite{Colomb}, in
the context of which products of distribution can (in principle) be handled.

\subsection{Embedding from higher dimensions}

The complete class of {\it nonexpanding} impulsive waves in spaces of constant
curvature with a {\it nonvanishing} cosmological constant~$\Lambda$ can
alternatively be introduced using a convenient 5-dimensional formalism as metrics
\begin{equation}
\d s^2=   - \d{Z_0}^2+ \d{Z_1}^2 + \d{Z_2}^2 +\d{Z_3}^2+\epsilon\,\d{Z_4}^2
  +\H(Z_2,Z_3,Z_4)\,\delta(Z_0+Z_1)\,(\d Z_0+\d Z_1)^2\ ,
\label{general}
\end{equation}
restricted by the  constraint (\ref{hyp}). Physically, this is
an embedding of impulsive {\it pp\,}-waves (\ref{general}) which propagate in
5-dimensional flat space onto the (anti-)de~Sitter hyperboloid.
The wave is absent when  $\H=0$, in which case (\ref{general}) with the
constraint (\ref{hyp}) reduces to the standard form of  the de~Sitter or
the anti-de~Sitter spacetime (\ref{ds0}). It has been demonstrated in \cite{[B6]}
that for a nontrivial $\H$ the metric represents impulsive waves propagating in the
(anti-)de~Sitter universe. Each impulse is located on the null hypersurface $
Z_0+Z_1=0$. Considering (\ref{hyp}), this is given by
 \begin{equation}
 {Z_2}^2+{Z_3}^2+\epsilon{Z_4}^2=\epsilon a^2\ ,\label{surface}
 \end{equation}
which is a nonexpanding 2-sphere in the de~Sitter universe or a hyperboloidal
2-surface in the anti-de~Sitter universe. The wave surfaces can naturally be
parametrized as
 \begin{equation}
Z_2=a\sqrt{\epsilon(1-z^2)}\,\cos\phi\ ,\quad
Z_3=a\sqrt{\epsilon(1-z^2)}\,\sin\phi\ ,\quad
Z_4=a\, z\ .  \label{param}
 \end{equation}

Various 4-dimensional parametrizations of the solutions
(\ref{general}) on (\ref{hyp}) can be considered. For example,
\begin{equation}
Z_0 = {\U+\V\over\sqrt2\,\Omega} \ , \qquad\   Z_1 = {\U-\V\over\sqrt2\,\Omega} \ ,
 \qquad Z_2+{\rm i}Z_3 = {\sqrt2\,\eta\over\Omega}\ ,  \qquad
 Z_4 = a \left({2\over\Omega}-1\right), \label{Zcoord}
 \end{equation}
where $\>\Omega=1+{\textstyle{1\over6}}\Lambda(\eta\bar\eta-\U\V)\, ,$
brings the metric to the previous form  (\ref{confppimp}) with
$\H=\sqrt2 H/{\textstyle(1+{1\over6}\,\Lambda\,\eta\, \bar\eta)}$.
Other natural coordinates which parametrize (\ref{general}) have been discussed
in \cite{[B5]}.

We have  remarked in previous sections that  the above metrics may describe
impulsive gravitational waves and/or impulses of null matter. Purely gravitational waves occur when
the   vacuum field equation
 \begin{equation}
(\Delta+{\textstyle{2\over3}}\Lambda)\,\H=0 \label{vacuum}
 \end{equation}
is satisfied \cite{HorItz99}, \cite{Sfet}, \cite{[B7]}, in which
$\Delta\equiv{1\over3}\Lambda\{\partial_z[(1-z^2)\partial_z]+(1-z^2)^{-1} \partial_\phi\partial_\phi)\}$
is the Laplacian on the impulsive surface.

\subsection{Boosts and limits of infinite acceleration}

Let us finally outline yet another fundamental method for the construction
of particular (but physically important) impulses in spaces
of constant curvature. It is based on boosting  suitable, initially
static sources to the speed of light which yields specific
nonexpanding impulsive waves. Similarly,  limits of infinite
acceleration of specific sources give special expanding impulsive solutions.

\subsubsection{ Nonexpanding impulsive waves}

It was first demonstrated in 1971 by Aichelburg and Sexl in a  classic paper
\cite{AS} that a specific impulsive gravitational {\it pp\,}-wave solution
can be obtained by boosting the Schwarzschild black hole to the speed of light,
while its mass is reduced to zero in an appropriate way. Let us start with the metric
 \begin{equation}
 \d s^2 =-\d t^2+\d r^2+r^2(\d\vartheta^2+\sin^2\vartheta\,\d\varphi^2)
+\Psi(\d t^2+ \d r^2)\ ,   \label{Schw}
 \end{equation}
 where  $\Psi=2M/r$, which is the Schwarzschild (static and spherically
 symmetric) solution linearized for small values of
 $M$. We introduce Cartesian coordinates $x=r\sin\vartheta\cos\varphi$,
$y=r\sin\vartheta\sin\varphi$, $z=r\cos\vartheta$, and  perform
a boost  in  the $x$-direction,
 \begin{equation}
 t={\tilde t+\vv\tilde x\over\sqrt{1-\vv^2}}\ , \qquad
x={\tilde x+\vv\tilde t\over\sqrt{1-\vv^2}}\ .\label{boost}
 \end{equation}
 In the limit as $\vv\to1$, the line element (\ref{Schw}) takes the  form
 \begin{equation}
 \d s^2 =-\d\tilde t^2+\d\tilde x^2+\d y^2+\d z^2
+2(\d\tilde t+\d\tilde x)^2 \lim_{\vv\to1}{\Psi\over1-\vv^2}\ ,\label{boo}
\end{equation}
where $\Psi=2M\,(x^2+y^2+z^2)^{-1/2}\,$ and $x$ is given by (\ref{boost}).
 To evaluate this limit, we employ the identity
 \begin{equation}
\lim_{\vv\to1} {1\over\sqrt{1-\vv^2}}\,\Psi(x) =\delta(\tilde t+\tilde x)
\int_{-\infty}^{+\infty} \Psi(x)\,\d x\ , \label{identity}
 \end{equation}
(see \cite{HotTan93}, \cite{[B5]})
 which is valid in the distributional sense. It is also necessary to scale
the parameter $M$ in $\Psi$ to zero   such that \
$8M=b_0\sqrt{1-\vv^2}\,$, where $b_0$ is a new constant. Obviously, we obtain an
impulsive {\it pp\,}-wave metric (\ref{imp-pp}),
 \begin{eqnarray}
 \d s^2 =-\d\tilde t^2+\d\tilde x^2+\d y^2+\d z^2
 +\H\,\delta(\tilde t+\tilde x)(\d\tilde t+\d\tilde x)^2\ ,
 \label{imppp}
 \end{eqnarray}
 where $ \H={1\over2}b_0\int_{-\infty}^{+\infty} (\rho^2+x^2)^{-1/2}\,\d x$,
and $\rho^2=y^2+z^2$. The divergence in the integral can be removed
 by   the transformation
$\tilde t-\tilde x \to\tilde t-\tilde x -{1\over2}b_0\,\lim_{\vv\to1}
\log\left(\tilde x+\vv\tilde t-\sqrt{(\tilde x+\vv\tilde
t)^2+1-\vv^2}\right)$,
 which gives
 \begin{equation}
    \H={\textstyle {1\over2}}b_0 \int_{-\infty}^{+\infty} \left[(\rho^2+x^2)^{-1/2}
  -(1+x^2)^{-1/2} \right]\,\d x =-b_0 \,\log\rho\ .\label{AiSe}
 \end{equation}
 This is the famous Aichelburg--Sexl solution \cite{AS} which represents an
axially-symmetric impulsive gravitational wave in Minkowski space
generated by a single null monopole particle located on $\rho=0$.
Using a similar approach, numbers of other specific impulsive
waves in flat space have been obtained by boosting various spacetimes of the
Kerr--Newman  \cite{KerNew} or the Weyl family \cite{[B4]}.

The method can be generalized to obtain impulses in spacetimes with $\Lambda\not=0$.
This was first done in 1993 by Hotta and Tanaka \cite{HotTan93} who
boosted the Schwarzschild--de~Sitter solution to obtain a nonexpanding spherical
impulsive gravitational wave generated by a pair of null monopole particles in
the de~Sitter background. They also described an analogous solution in the
anti-de~Sitter universe. Their main ``trick'' was to consider the boost in the 5-dimensional
representation of the (anti-)de~Sitter spacetime (\ref{hyp}), (\ref{ds0}) where the
boost can explicitly (and consistently) be performed. One starts with the line
element
 \begin{eqnarray}
 \d s^2 &=&-\left(1-\epsilon\,{r^2/a^2}\right)\d t^2
+\left(1-\epsilon\,{r^2/a^2}\right)^{-1}\d r^2 \nonumber +r^2(\d\vartheta^2+\sin^2\vartheta\d\varphi^2) \nonumber \\
&&\hskip2cm +\Psi\left[ \d t^2+ \left(1-\epsilon\,{r^2/ a^2}\right)^{-2}\d r^2 \right],
\label{schds}
 \end{eqnarray}
 where  $\Psi=2M/r$. This is a perturbation of the de~Sitter ($\epsilon=+1$) or
 anti-de~Sitter ($\epsilon=-1$) space corresponding to the
Schwarzschild--(anti-)de~Sitter solution \cite{SdS}, linearized for small $M$.
As $\Lambda\to0$, the line element (\ref{schds}) reduces to  the
Schwarzschild-like perturbations of the Minkowski spacetime (\ref{Schw}).
For $M=0$ the metric (\ref{schds}) is the (anti-)de~Sitter spacetime in
standard ``static'' coordinates. These parametrize the hyperboloid (\ref{hyp})
by $Z_1=r\sin\vartheta\cos\varphi$,
$Z_3=r\sin\vartheta\sin\varphi$,
$Z_2=r\cos\vartheta$,
 and
 \begin{eqnarray}
\hbox{for } \epsilon=+1:&&
Z_0=\sqrt{a^2-r^2}\,\sinh(t/a)\ , \quad
 Z_4=\pm\sqrt{a^2-r^2}\,\cosh(t/a)\  ,\nonumber\\
\hbox{for } \epsilon=-1:&&
Z_0=\sqrt{a^2+r^2}\,\sin(t/a)\ , \quad\ \>
 Z_4=\sqrt{a^2+r^2}\,\cos(t/a)\ . \label{static}
  \end{eqnarray}
 Writing (\ref{schds}) as $\>\d s^2=\d s_0^2+\d s_1^2$, where $\>\d s_0^2$
 has the form (\ref{ds0}),  we may now express the perturbation  as
 \begin{equation}
 \d s_1^2 =a^2 \Psi  \Bigg[
\left( {Z_4\d Z_0-Z_0\d Z_4\over{Z_4}^2-\epsilon{Z_0}^2} \right)^2
+{a^2\over r^2} \left( {Z_0\d Z_0-\epsilon Z_4\d Z_4\over{Z_4}^2-\epsilon{Z_0}^2} \right)^2
\Bigg]\ , \label{pertZ}
 \end{equation}
 where $r=\sqrt{{Z_0}^2+\epsilon(a^2-{Z_4}^2)}$.
Performing a boost in the $Z_1$-direction,
 \begin{equation}
Z_0={\tilde Z_0+\vv\tilde Z_1\over\sqrt{1-\vv^2}}\ , \qquad
Z_1={\tilde Z_1+\vv\tilde Z_0\over\sqrt{1-\vv^2}}\ ,
 \end{equation}
 the (anti-)de~Sitter background $\>\d s_0^2$ is, of course, invariant.
Using again the identity (\ref{identity}) and rescaling the coefficient
as $8M=b_0\sqrt{1-\vv^2}$, in the limit $\vv\to1$ the term $\d s_1^2$ takes the form
 \begin{equation}
  \d s_1^2 =\H\,\delta(\tilde Z_0+\tilde Z_1)\, (\d\tilde Z_0+\d\tilde Z_1)^2\ ,\label{ds1}
 \end{equation}
 where
 \begin{equation}
 \H(Z_4) ={\textstyle{1\over4}}b_0\, a^2
   \int_{-\infty}^{+\infty} {\epsilon(a^2-{Z_4}^2){Z_4}^2+(a^2+{Z_4}^2)x^2 \over
\left({Z_4}^2-\epsilon x^2\right)^2 \left[x^2+\epsilon(a^2-{Z_4}^2)\right]^{3/2}}\, \d x\ .
\label{integral}
 \end{equation}
We immediately observe that the structure of the boosted metric is exactly that
of (\ref{general}) with the impulse located on  (\ref{surface}).
Considering the parametrization (\ref{param}) of this wave surface
and substituting $\theta$ for $x$ so that
$x=a\sqrt{\epsilon(1-z^2)}\cot\theta$, we explicitly evaluate the
integral (\ref{integral}) as
 \begin{equation}
\H={\textstyle{1\over4}}b_0\int_0^\pi {z^2+\cos^2\theta\over(z^2-\cos^2\theta)^2}
\, \sin^3\theta \,\d\theta
=b_0\left(\,{z\over2}\log\left|{1+z\over1-z}\right|-1\,\right)\ . \label{HTimp}
 \end{equation}
 This is the axially-symmetric Hotta--Tanaka solution \cite{HotTan93}.
Further details on  boosting  monopole (and also multipole) particles
to the speed of light in the (anti-)de~Sitter universe, the geometry of
the nonexpanding wave surfaces, and discussion of
various useful coordinates can be found in~\cite{[B5]}, \cite{[B4]}.
In particular, it was demonstrated that although the impulsive
wave surface is nonexpanding, for $\Lambda>0$ this coincides with
the horizon of the closed de~Sitter universe. The background space
contracts to a minimum size and then reexpand in such a way that
the nonexpanding impulsive wave in fact propagates with the speed
of light from the ``north pole'' of the universe across the
equator to its ``south pole''.

\subsubsection{ Expanding impulsive waves}

As shown above, nonexpanding  impulsive waves can be obtained by boosting
particular (initially static) sources to the speed of light.
Interestingly, specific expanding impulsive spherical gravitational
waves can also be obtained by considering exact solutions representing
accelerating sources and taking these to the limit in which the
acceleration becomes unbounded. It was first realized by Ji\v r\'\i \ Bi\v c\'ak
\cite{Bi90} (see also \cite{BiSc89}, with recent generalizations in \cite{[B11]},
\cite{[B12]}) that expanding impulses can be obtained from boost-rotation symmetric
solutions which represent  gravitational field of uniformly accelerating objects.
In the limit of infinite acceleration, these well-known  explicit exact
radiative spacetimes form spherical impulsive waves  attached to conical
singularities.

As Bi\v c\'ak writes in his essay \cite{JiBi2000},
``the boost-rotation symmetric solutions representing uniformly
accelerated objects ... play
a unique role among radiative spacetimes since they are
asymptotically flat, in the sense that they admit global smooth
sections of null infinity. And as the only known radiative
solutions describing finite sources they can provide expressions
for the Bondi mass, the news function, or the radiation patterns
in explicit forms.'' Special solutions of this type have been
studied for almost forty years, first by Bondi \cite{Bondi57},
Bonnor and Swaminarayan \cite{BonSwa64},  Israel and Khan \cite{IKH},
and then by many others --- but Ji\v r\'\i \  Bi\v{c}\'ak
in particular. In fact, the investigation of various aspects of radiative
spacetimes and fields with boost-rotation symmetry is his ``most typical''
topic. The analysis of radiative
properties of the Bonnor--Swaminarayan solutions became the basis
of Bi\v{c}\'ak's Ph.\,D. thesis and of his first paper published abroad
\cite{Bic68}. He realized that these belong to a wide class of
boost-rotation symmetric solutions (of which, for example, the
well-known $C$-metric is also a member) and that the boost
symmetry is the {\it only} admissible second symmetry of axially
symmetric, asymptotically flat spacetimes which admit gravitational
radiation. He and his collaborators have studied the unique properties
of these spacetimes, including their global structure,  in a number
of fundamental works, e.g.  \cite{Bic68}-\cite{BicPra99},
\cite{Bi90}, \cite{BiSc89}. He has also summarized these results in many
inspiring reviews \cite{Bic71}-\cite{Bic00}, \cite{JiBi2000}. Altogether,
the contributions concerning boost-rotation symmetric solutions represent almost
a quarter of all Bi\v c\'ak's original publications.

The boost-rotation symmetric  spacetimes can be described by the line element
 \begin{equation}
 \d s^2=e^\lambda \d\rho^2 + \rho^2 e^{-\mu} \d \phi^2 +
  (\zeta^2-\tau^2)^{-1} \left[e^\lambda(\zeta \d \zeta - \tau \d\tau)^2
  -e^\mu (\zeta\d\tau-\tau\d\zeta)^2\right]\ ,
 \label{BSmetric}
 \end{equation}
where specific metric functions $\mu$ and $\lambda$ depend only on
$\zeta^2-\tau^2$ and $\rho^2$. In particular, the Bonnor--Swaminarayan
solution \cite{BonSwa64} generally contains five arbitrary constants $m_1$, $m_2$,
$A_1$, $A_2$, and $B$ which determine the masses and uniform accelerations
of two pairs of particles, and the singularity structure on the axis
of symmetry $\rho=0$ (see e.g. \cite{BonSwa64}, \cite{Bic85}, \cite{[B11]}).
It is possible to choose the
constant $B$  such that some sections of the axis are regular. Of particular
interest are special cases described by Bi\v{c}\'ak, Hoenselaers and Schmidt
\cite{BHS1}, \cite{BHS2} which represent only {\it two} (Curzon--Chazy)
particles\footnote{In \cite{[B11]} we investigated systematically all the
spacetimes with four accelerating particles. Their  limits $A_i\to\infty$
are unphysical,  with the only  exception of a particular solution which
represents a finite string/strut between the two outer particles which snaps/breaks
at its midpoint and the two broken ends separate at the speed of light.}
of mass $m$ which are accelerating in opposite directions with
acceleration $A$. For these solutions the metric functions
in (\ref{BSmetric}) can be written in a simple form
 \begin{eqnarray}
  \mu&=&-{2m\over A R}+  4m A+ B\ , \nonumber \\
  \lambda&=&-{m^2\over A^2R^4}\,\rho^2(\zeta^2-\tau^2)
  +{2m A\over R}\,(\rho^2 + \zeta^2 - \tau^2) + B\ ,   \label{BSmono}
\end{eqnarray}
where $ R={1\over 2}\sqrt{\left(\rho^2+\zeta^2-\tau^2- A^{-2}\right)^2+ 4 A^{-2}\rho^2}$.
For $B=0$, the axis $\rho=0$ is  regular between the symmetrically located particles
but these are connected to infinity by two semi-infinite strings which cause
their acceleration. For $B=-4mA$ there is a finite expanding strut  between the
particles.

The  limit of infinite acceleration $A\to\infty$ of these solutions
was investigated by Bi\v{c}\'ak  \cite{Bi90}, \cite{BiSc89}.
It is necessary to scale  the mass parameter $m$  to zero in such a way
that the ``monopole moment'' $M_0=-4mA$ remains constant. In this
limit we obtain
\begin{eqnarray}
\mu&=& B-M_0\> , \nonumber\\
\lambda&=& B-{\rm sign}\,(\,\rho^2+\zeta^2-\tau^2\,)\,M_0\ . \label{mulambda}
\end{eqnarray}
The resulting spacetime is locally flat
everywhere except on the sphere $\rho^2+\zeta^2=\tau^2$.
It therefore describes an expanding spherical impulsive
gravitational wave generated by two particles which move apart
at the speed of light in the Minkowski background and are connected to infinity by
two semi-infinite strings ($B=0$), or each other by an expanding strut ($B=M_0$).
Performing the transformation (see, e.g. \cite{BicSch89b})
$\rho=\textstyle{1\over2}(v-u)$,
$\tau =\pm\textstyle{1\over2}(v+u)\,\cosh\chi$,
$\zeta=\textstyle{1\over2}(v+u)\,\sinh\chi$,
 we  put the solution (\ref{BSmetric}), (\ref{mulambda}) into the standard form of
boost-rotational symmetric spacetimes
\begin{equation}
 \d s^2= \textstyle{1\over4}(v-u)^2 \,e^{-\mu} \,\d\phi^2
 + \textstyle{1\over4}(v+u)^2 \,e^\mu \,\d \chi^2 -e^\lambda \,\d u\,\d v \ .
 \label{BSnull}
\end{equation}
The metric functions are
$\mu= -M_0$, $\lambda=\left[\,\Theta(uv)-\Theta(-uv)\,\right]M_0$\
for the two receding strings,
or $\mu= 0$, $\lambda=2\,\Theta(uv)M_0$\  for the expanding strut.
In both cases $\mu$ is a constant, but there is a discontinuity in the otherwise
constant value of $\lambda$ with the step \ $2M_0$ \ on the null cone \ $uv=0$. \
It is possible to perform a further transformation to coordinates in
which the metric is  continuous everywhere.
Indeed, in the region where the functions $\mu$ and $\lambda$ are
constant, the transformation
$U= -u\,e^{\lambda+\mu/2}$,
$V= {\textstyle{1\over2}}\,v\,e^{-\mu/2}$,
$\psi= \chi\,e^\mu$,
brings (\ref{BSnull}) into the form
 \begin{equation}
 \d s^2=\big(V+{\textstyle{1\over2}}P\, U \big)^2 \,\d \phi^2
 +\big(V-{\textstyle{1\over2}}P\, U \big)^2 \,\d \psi^2 +2 \,\d  U\, \d V\  ,
 \label{GP}
 \end{equation}
where $P=e^{-(\lambda+\mu)}$. The solution for {\it two receding strings} can thus be written in the
form (\ref{GP}) with
\begin{equation}
 P=\Theta(- U)+\beta^2\, \Theta( U)    \  ,
 \label{ABsnap}
\end{equation}
where $\beta=\exp (M_0)$.  The metric (\ref{GP}), (\ref{ABsnap}) is exactly
that constructed by Gleiser and Pullin \cite{GlePul89} by their ``cut and paste''
method. It represents an impulsive spherical gravitational wave  propagating in the
Minkowski universe. Outside the wave ($U>0$), there
are two receding strings characterized by a deficit angle $(1-\beta)2\pi$,
which can be interpreted as remnants of one cosmic string which ``snapped".
However, as pointed out by Bi\v{c}\'ak
\cite{Bi90}, \cite{JiBiunpub}, the complete solution rather describes two semi-infinite
strings approaching at the speed of light and separating again at the
instant at which they collide.
The complementary solution for an {\it expanding strut} is also given by (\ref{GP}) with
\begin{equation}
 P= \beta^{-2}\,\Theta(-U) +\Theta(U)\ .
 \label{ABexp}
\end{equation}
In this case there is a string with the deficit angle $(1-\beta^{-1})2\pi$
in the locally Minkowski space inside the  impulsive wave ($U<0$).

Recall that there is an alternative continuous form of the above
solutions. Introducing  $Z={1\over\sqrt2}(\psi+{\rm i}\,\phi)$,
we  convert the metric (\ref{GP}), (\ref{ABsnap}) into the form
$ \d s^2=2\,|\,V \d Z + U H \d \bar Z\,|^2 + 2\,\d  U \,\d  V$,
where  $H=-{1\over2}\beta^2$. Performing another transformation of
the metric (\ref{BSnull}) in the region  $u>0$, $v>0$ as
\begin{eqnarray}
 U &=& -\> {uv\over u+v}\,\exp
  \left[ {\textstyle{\lambda\over2}} -\chi\,e^{(\mu-\lambda)/2} \
\right]
\>,  \nonumber\\
 V &=& {\textstyle{1\over2}}\,(u+v)\,\exp
  \left[ {\textstyle{\lambda\over2}}+\chi\,e^{(\mu-\lambda)/2} \ \right]
\>,  \label{trans2b}\\
 Z &=& {\textstyle{1\over\sqrt2}}\,{v-u\over v+u}\,\exp\left[
  -\chi\,e^{(\mu-\lambda)/2}
  + {\rm i}\,\phi\,e^{-(\mu+\lambda)/2} \ \right] \>,  \nonumber
\end{eqnarray}
 we
obtain $\d s^2=2\,V^2 \,\d Z \,\d \bar Z + 2\,\d  U \,\d  V$.
These two metrics for $U>0$ and $U<0$ can be matched continuously
across the null cone  $U=0$. In fact, this is exactly a particular
case of the metric  (\ref{en0}) for $\Lambda=0$, $\epsilon=0$, and
constant $H=-{1\over2}\beta^2$, which describes impulsive spherical
wave \cite{Pen72}, \cite{NutPen92}, \cite{Hogan93}, \cite{[B8]}.

Recently  we analyzed  \cite{[B11]} the limit of infinite acceleration of
much larger class of  boost-rotationally symmetric spacetimes
which generalize the Bonnor--Swaminarayan solution. These explicit solutions
were found by Bi\v{c}\'ak, Hoenselaers and Schmidt  \cite{BHS2}
and represent  two uniformly accelerating particles with an {\it arbitrary multipole
structure} attached to conical singularities (as in the two cases  above).
These  solutions can be written as \cite{[B11]}
\begin{eqnarray}
 \mu&=& 2\sum_{n=0}^{\infty}M_n \, {P_n\over{(x-y)^{n+1}} } + B-M\ ,
\nonumber\\
\lambda&=&-2\sum_{k,l=0}^{\infty} M_k M_l \,{(k+1)(l+1)\over(k+l+2)}\,
   {(P_kP_l-P_{k+1}P_{l+1})\over {(x-y)^{k+l+2}}} \label{lambdamulti}\\
   &&\qquad\qquad -\left({x+y\over x-y}\right)
\sum_{n=0}^{\infty} {M_n\over 2^n}\,\sum_{l=0}^n
\left({2\over x-y}\right)^l P_l + B\ ,
\nonumber
 \end{eqnarray}
where the constants $M_n$ represent the multipole moments,
the argument of the Legendre polynomials $P_n$ is $\alpha=(1-xy)/(x-y)$,
 $B$ is a constant,
\begin{equation}
M\equiv\sum_{n=0}^{\infty}\ {M_n\over 2^n}\ ,\label{M}
\end{equation}
and for $x$, $y$ and  $\alpha$ in (\ref{lambdamulti}) one has to substitute from
\begin{equation}
x-y=4A^2R\ , \qquad x+y=2A^2(\rho^2+\zeta^2-\tau^2)\ ,\qquad
\alpha={\textstyle{1\over 2}}(\rho^2-\zeta^2+\tau^2+A^{-2})/R\ .
 \label{alpha}
\end{equation}
The axis is regular {\it between} the two particles if  $B=0$.
In this case there are strings connecting the particles to
infinity. The alternative situation with a strut between the particles
(and the axis regular {\it outside}) is given by $B=M$.
The axis is  regular {\it everywhere} (except at the particles) if
the combination of multipole moments satisfies the condition
$M=0$ and we assume  $B=0$. In this case the particles are self-accelerating
\cite{BHS2}. Considering only the case $n=0$ in (\ref{lambdamulti}), we recover the
previous solution (\ref{BSmono}) for the accelerating monopole
particles, with $M=M_0=-4mA$.

Now, we consider the null limit $A\to\infty$ of the general class of solutions
(\ref{lambdamulti}) in  which all the multipole moments $M_n$ are kept constant.
Interestingly, in this limit we again obtain  (\ref{mulambda}), only the parameter $M_0$
is now replaced by a more general parameter $M$ introduced in (\ref{M}).
For the particular values of the constant $B$ described above
this solution describes a snapping cosmic string,
or an expanding strut. The ends of the strings/strut move in opposite directions
with the speed of light,  generating an impulsive spherical gravitational
wave. In the limit, the multipole structure of  initial particles {\it disappears}
and the solution is characterised  by the single constant $M$ only.
Thus, the  limit $A\to\infty$ of any uniformly accelerating
multipole particle  is identical to that of a monopole particle $M_0=M$,
as obtained originally in \cite{Bi90}, \cite{BiSc89}. Of course, the metric can
also be represented in the coordinates  (\ref{GP}) or (\ref{en0}) with
$H=-{1\over2}\beta^2$, the only difference is that the parameter $\beta$ has
a more general form
\begin{equation}
\beta=\exp\left(\sum_{n=0}^{\infty}\>{M_n\over 2^n}\right)\ .\label{beta}
\end{equation}

In the subsequent paper \cite{[B12]} we also investigated all possible
null limits $A\to\infty$ of another important explicit class of
solutions with boost-rotation symmetry, namely the well-know
$C$-metric
 \begin{equation}
 \d s^2 = -A^{-2}(x+y)^{-2} \left(F^{-1}\d y^2 + G^{-1}\d x^2 + G\d\phi^2
-F\d t^2 \right),
 \label{Cmetric}
 \end{equation}
where $F = -1 + y^2 - 2m Ay^3$, $G = 1 - x^2 - 2m Ax^3$. It was
shown already in 1970 by Kinnersley and Walker \cite{KinWal70} that this
metric represents two uniformly accelerating {\it black holes}, each of mass $m$.
The acceleration $A$ is caused either by a strut between
the black holes or by two semi-infinite strings connecting them to
infinity. Radiative and asymptotic properties were investigated in
\cite{FarZim80}. Bonnor \cite{Bon83} found an explicit transformation
of (\ref{Cmetric}) into a form (\ref{BSmetric}). When
$mA<1/(3\sqrt3)$ there are four possible spacetimes according to different
ranges of the coordinates $x$ and $y$ \cite{CorUtt95}, and the
corresponding explicit functions $\lambda$, $\mu$  have more complicated
forms. In \cite{[B12]} we investigated  the limits  $A\to\infty$ of all these
possibilities. It was demonstrated that (scaling again the mass parameter $m$
to zero such that $mA$ remains constant) this limit is {\it identical}
to the metric  (\ref{GP}) of a spherical  impulsive
gravitational wave generated either by a snapping string
(\ref{ABsnap}), or an expanding strut (\ref{ABexp}), with
 \begin{equation}
\beta= \textstyle{{1\over2}
\,\big[\,1+\sqrt3\cot(\varphi+{1\over3}\pi)\,\big]}\>\in(0,1]\ ,\quad
\hbox{\ where\ \ } \varphi={1\over3}\arccos(1-54\,m^2A^2) \ . \label{betaC}
 \end{equation}

Note finally that it is natural to expect that the analogous null limit
of infinite acceleration $A\to\infty$ of a more general $C$-metric, which admits
a nonvanishing value of the cosmological constant $\Lambda\>$ \cite{PleDem76},
would generate an expanding spherical impulsive wave (\ref{en0})
in the (anti-)de~Sitter universe. However, such limit is
mathematically  more involved and has not yet been explicitly performed.
We are currently investigating this problem. A first step has been achieved
in \cite{[B13]} where a physical interpretation of the parameters of the solutions
\cite{PleDem76}, which represent uniformly accelerating black holes in spacetimes of constant
nonvanishing curvature, was presented (see also \cite{ManRos95}).
However, a deeper understanding of the global structure of these solutions
is required. The most recent contribution by Bi\v{c}\'ak and
Krtou\v{s} \cite{BicKr} may be of a great help.

\newpage

\subsection{Summary}

Let us now  summarize for convenience all the methods described
above for the construction of impulsive waves in spacetimes of
constant curvature, together with the main corresponding
references. This is presented in the following four diagrams,
which separately describe the construction of nonexpanding/expanding
waves in Minkowski/(anti-)de~Sitter universes.

\subsubsection{ Nonexpanding impulsive waves}

\vskip10mm
\

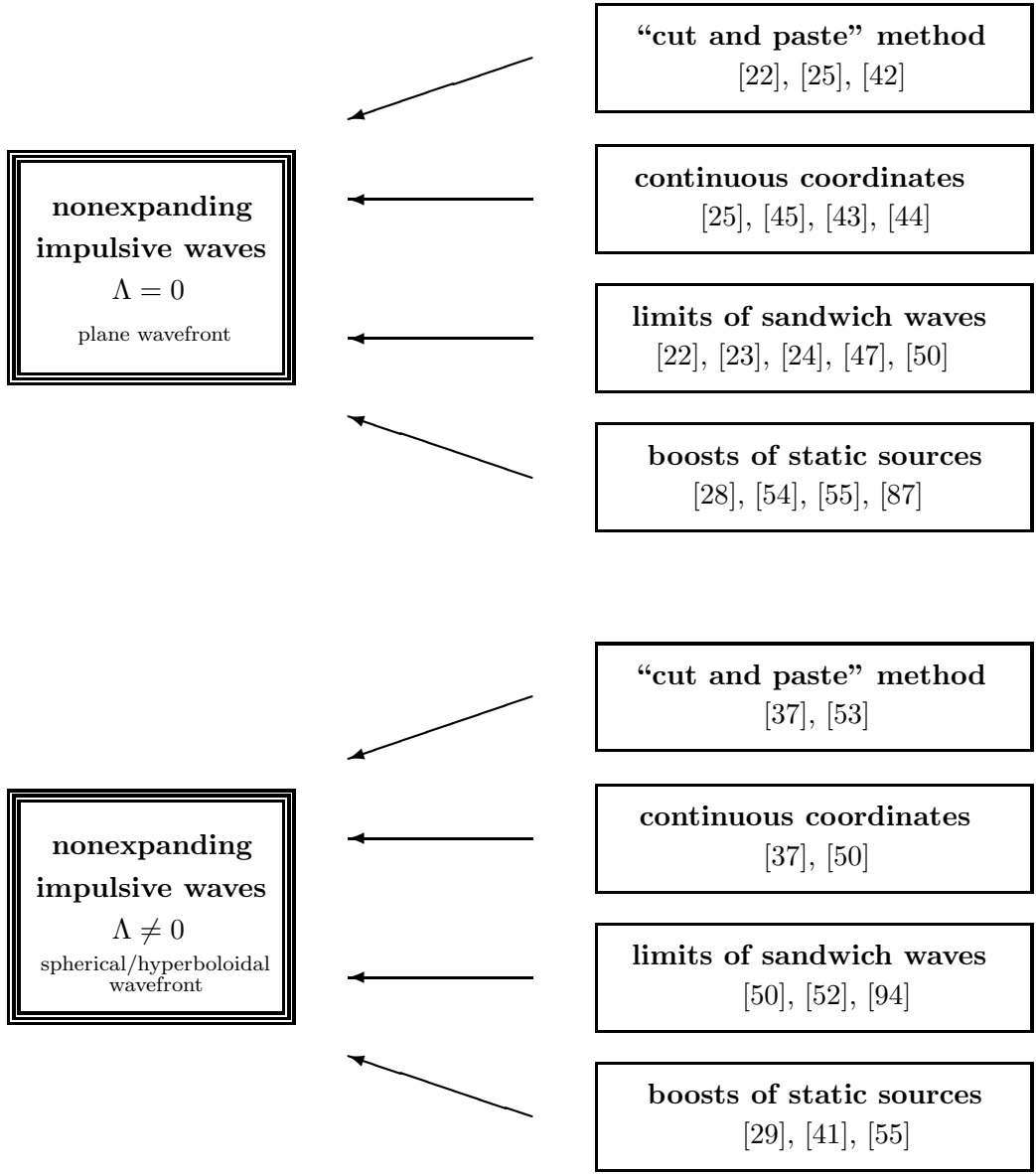
\begin{picture}(450,200)
{\thicklines
\put(20,60){\framebox(100,80){}}
\put(18,58){\framebox(104,84){}}
\put(16,56){\framebox(108,88){}}
\put(32,120){{\bf nonexpanding}}
\put(26,104){{\bf impulsive waves}}
\put(55,88){$\Lambda=0$}
\put(42,72){{\scriptsize plane wavefront}}
\put(240,160){\framebox(165,40){}}
\put(254,185){{\bf ``cut and paste'' method}}
\put(293,170){\cite{Pen68}, \cite{Pen72}, \cite{DaT}}
\put(240,106){\framebox(165,40){}}
\put(254,131){{\bf continuous coordinates}}
\put(279,116){\cite{Pen72}, \cite{DE78}, \cite{[B2]}, \cite{Steinb}}
\put(240,53){\framebox(165,40){}}
\put(253,78){{\bf limits of sandwich waves}}
\put(262,63){\cite{Pen68}, \cite{Rindler}, \cite{[A3]},  \cite{Steina}, \cite{[B1]}}
\put(240,0){\framebox(165,40){}}
\put(259,25){{\bf boosts of static  sources}}
\put(276,10){\cite{AS}, \cite{KerNew}, \cite{[B4]}, \cite{[B3]}}
\put(215,180){\vector(-3,-1){70}}
\put(215,126){\vector(-1, 0){70}}
\put(215, 73){\vector(-1, 0){70}}
\put(215, 20){\vector(-3, 1){70}}
}
\end{picture}
\vskip15mm

\noindent

\begin{picture}(450,200)
{\thicklines
\put(20,60){\framebox(100,80){}}
\put(18,58){\framebox(104,84){}}
\put(16,56){\framebox(108,88){}}
\put(32,120){{\bf nonexpanding}}
\put(26,104){{\bf impulsive waves}}
\put(55,88){$\Lambda\not=0$}
\put(28,76){{\scriptsize spherical/hyperboloidal}}
\put(54,68){{\scriptsize wavefront}}
\put(240,160){\framebox(165,40){}}
\put(254,185){{\bf ``cut and paste'' method}}
\put(303,170){\cite{[B7]}, \cite{Sfet}}
\put(240,106){\framebox(165,40){}}
\put(256,131){{\bf continuous coordinates}}
\put(303,116){\cite{[B7]}, \cite{[B1]}}
\put(240,53){\framebox(165,40){}}
\put(253,78){{\bf limits of sandwich waves}}
\put(295,63){\cite{[B1]}, \cite{[B6]}, \cite{Defrise}}
\put(240,0){\framebox(165,40){}}
\put(259,25){{\bf boosts of  static  sources}}
\put(295,10){\cite{HotTan93}, \cite{[B5]}, \cite{[B4]}}
\put(215,180){\vector(-3,-1){70}}
\put(215,126){\vector(-1, 0){70}}
\put(215, 73){\vector(-1, 0){70}}
\put(215, 20){\vector(-3, 1){70}}
}
\label{schema}
\end{picture}
\vskip10mm

\noindent
In addition, nonexpanding impulses with $\Lambda\ne0$ can also be  constructed by
embedding  impulsive {\it pp\,}-waves from higher dimensions on the
(anti-)de~Sitter hyperboloid, see the references
\cite{HotTan93}, \cite{[B5]},  \cite{HorItz99}, \cite{[B6]}, \cite{[B7]}.

\newpage

\subsubsection{ Expanding impulsive waves}

\vskip10mm
\

\begin{picture}(450,200)
{\thicklines
\put(20,60){\framebox(100,80){}}
\put(18,58){\framebox(104,84){}}
\put(16,56){\framebox(108,88){}}
\put(40,120){{\bf expanding}}
\put(24,104){{\bf impulsive waves}}
\put(55,88){$\Lambda=0$}
\put(36,72){{\scriptsize spherical wavefront}}
\put(238,160){\framebox(165,40){}}
\put(254,185){{\bf ``cut and paste'' method}}
\put(255,170){\cite{Pen72}, \cite{GlePul89}, \cite{NutPen92},
  \cite{Hogan93}, \cite{Hogan94}, \cite{[B10]}}
\put(238,106){\framebox(165,40){}}
\put(256,131){{\bf continuous coordinates}}
\put(255,116){\cite{NutPen92}, \cite{Hogan93},
  \cite{Hogan94}, \cite{[B8]}, \cite{[B10]}, \cite{NutAli}}
\put(238,53){\framebox(165,40){}}
\put(243,78){{\bf limits of sandwich RT waves}}
\put(297,63){ \cite{[B8]}, \cite{[B9]}}
\put(238,0){\framebox(165,40){}}
\put(242,25){{\bf infinitely accelerated sources}}
\put(273,10){  \cite{Bi90}, \cite{BiSc89}, \cite{[B11]}, \cite{[B12]}}
\put(215,180){\vector(-3,-1){70}}
\put(215,126){\vector(-1, 0){70}}
\put(215, 73){\vector(-1, 0){70}}
\put(215, 20){\vector(-3, 1){70}}
}
\end{picture}
\vskip15mm

\begin{picture}(450,200)
{\thicklines
\put(20,60){\framebox(100,80){}}
\put(18,58){\framebox(104,84){}}
\put(16,56){\framebox(108,88){}}
\put(40,120){{\bf expanding}}
\put(24,104){{\bf impulsive waves}}
\put(55,88){$\Lambda\not=0$}
\put(36,72){{\scriptsize spherical wavefront}}
\put(238,160){\framebox(165,40){}}
\put(254,185){{\bf ``cut and paste'' method}}
\put(285,170){\cite{Hogan92}, \cite{[B8]}, \cite{[B10]}}
\put(238,106){\framebox(165,40){}}
\put(256,131){{\bf continuous coordinates}}
\put(285,116){\cite{Hogan92}, \cite{[B8]}, \cite{[B10]}}
\put(238,53){\framebox(165,40){}}
\put(243,78){{\bf limits of sandwich RT waves}}
\put(315,63){\cite{[B8]}}
\put(238,0){\framebox(165,40){}}
\put(242,25){{\bf infinitely accelerated sources}}
\put(320,10){---}
\put(215,180){\vector(-3,-1){70}}
\put(215,126){\vector(-1, 0){70}}
\put(215, 73){\vector(-1, 0){70}}
\put(215, 20){\vector(-3, 1){70}}
}
\end{picture}
\vskip10mm

\section{Particular solutions and some other properties}

In the remaining part of this review  some other properties of
impulsive waves in spaces of constant curvature are briefly described.
First, we present  particular solutions of this type which are of
interest from a physical point of view. Subsequently,  we  characterize
geodesics and symmetries in these spacetimes.

\subsection{Nonexpanding impulses generated by null multipole particles}

It has been demonstrated above in section 3.5.1 that the simplest
nonexpanding impulsive gravitational waves can be
generated by boosting  the Schwarzschild--(anti-)de~Sitter black hole
solutions. In the case $\Lambda=0$ this gives the Aichelburg--Sexl solution
(\ref{imppp}), (\ref{AiSe}), and for $\Lambda\not=0$ one obtains the
Hotta--Tanaka solution (\ref{general}), (\ref{HTimp}). Thus, the simplest
impulsive solutions can be regarded as  limits of
static ``monopole'' solutions boosted to the speed of light. This result can
be generalized:  it has been shown \cite{[B4]} that it is
possible to  consider particular nonexpanding impulsive waves generated by
null {\it multipole} particles as limits of boosted static multipole particles.

It is well-known \cite{KSMH}, \cite{Que} that   static, axially symmetric and
asymptotically flat vacuum solutions which represent  external field of sources
with a multipole structure can be written in the Weyl coordinates as
 \begin{equation}
 \d s^2=-e^{2\psi}\d t^2 + e^{-2\psi} \big[e^{2\gamma}(\d\varrho^2+\d z^2)
+\varrho^2\d\varphi^2 \big]\ ,\label{Weyl}
 \end{equation}
where
 \begin{eqnarray}
 \psi&=&\sum_{m=0}^\infty {a_m\over r^{m+1}}\,P_m(\cos\theta)\ , \\
\gamma&=&\sum_{m,n=0}^\infty {(m+1)(n+1)\over m+n+2}
{a_ma_n\over r^{m+n+2}} \big( P_{m+1}P_{n+1}-P_mP_n\big)\ , \nonumber
 \end{eqnarray}
$r=\sqrt{\varrho^2+z^2}$, $\cos\theta=z/r$, and $P_m$ are the
Legendre polynomials with argument $\cos\theta$. Arbitrary
constants $a_m$ determine the $m^{\rm th}$ multipole moments of the source.
Performing the boost (\ref{boost}) where $x=\varrho\cos\varphi$,
$y=\varrho\sin\varphi$, the  line element (\ref{Weyl})
in the limit $\vv\to1$ may be written in the
form (\ref{boo}) with $\Psi=-2\psi$. Considering again the
identity (\ref{identity}) and  rescaling
the parameters $a_m$ to zero   such that
$8a_m=-m \,b_m\sqrt{1-\vv^2}\,$, with $b_m$ being new constants which characterize
the corresponding multipole moments of the boosted source, we obtain the impulsive
{\it pp\,}-wave metric (\ref{imppp}) where
 \begin{equation}
 \H=\sum_{m=0}^\infty b_m \H_m
=\sum_{m=0}^\infty {\textstyle{1\over2}}m\, b_m\,\int_{-\infty}^{+\infty} {1\over(\rho^2+x^2)^{(m+1)/2}}\,
P_m\!\left({z\over\sqrt{\rho^2+x^2}} \right) \d x\ . \label{integra}
 \end{equation}
 For the simplest case $a_0\ne0$, $a_m=0$ for $m\ge1$ (which
corresponds to the boosted Curzon--Chazy solution) the integral
gives exactly the same result (\ref{AiSe}) as the Aichelburg--Sexl solution
\cite{AS} generated by a single null monopole particle.
For the higher multipole components  $m\ge1$ the integral
(\ref{integra}) can also be explicitly evaluated. As shown in \cite{[B4]}, we obtain
 \begin{equation}
 \H_m(\rho,\phi) =\rho^{-m}\cos [m(\phi-\phi_m)]\ ,
 \end{equation}
 where $\rho^2=y^2+z^2$, $\cos(\phi-\phi_m)=z/\rho$, and $\phi_m$ is a constant.
 This  term  represents the $m^{\rm th}$ multipole component
of the exact impulsive purely gravitational {\it pp\,}-wave
generated by a source of an arbitrary multipole structure \cite{[B3]}.
Indeed, the field equations for the metric (\ref{imppp}) with
 \begin{equation}
\H= -b_0\log\rho +\sum_{m=1}^\infty b_m\,\rho^{-m}\cos[m(\phi-\phi_m)]\ ,
 \end{equation}
correspond to a source localized  on the impulsive wavefront
$u\equiv\tilde t+\tilde x=0$ at $\rho=0$,
 which is described by $T_{uu}=J(\rho,\phi)\,\delta(u)$ where
 \begin{equation}
J(\rho,\phi)={\textstyle{1\over8}}b_0\,\delta(\rho) + \sum_{m=1}^\infty
{\textstyle{1\over8}}b_m{(-1)^m\over(m-1)!}\,\delta^{(m)}(\rho)\cos[m(\phi-\phi_m)]\ .
 \end{equation}
Clearly, the parameters $b_m$ and $\phi_m$  represent the
amplitude and phase of each multipole component.

Interestingly, there are analogous  impulsive solutions also
in the case $\Lambda\not=0$. It was demonstrated in  \cite{[B6]} that nontrivial
solutions of the vacuum Einstein equation (\ref{vacuum}) for the metric (\ref{general})
 on (\ref{hyp}), expressed in terms of the parameters (\ref{param}), can be written as
 \begin{equation}
\H(z,\phi)= \sum_{m=0}^\infty b_m\H_m
= \sum_{m=0}^\infty b_mQ^m_1(z)\cos[m(\phi-\phi_m)]\ , \label{E4.5}
\end{equation}
 where  $Q^m_1(z)$ are associated Legendre functions of the second kind generated
by the relation
 \begin{equation}
Q^m_1(z)=(-\epsilon)^m|1-z^2|^{m/2}{\d^mQ_1(z)\over \d z^m}\ .\label{Leg}
 \end{equation}
The first term for $m=0$,
 \begin{equation}
Q_1(z)\equiv Q^0_1(z)={z\over2}\log\left|1+z\over 1-z\right|-1\ ,  \label{HT}
 \end{equation}
exactly corresponds to the simplest axisymmetric Hotta--Tanaka solution (\ref{HTimp}).
The higher  components $\H_m$ describe nonexpanding impulsive gravitational
waves in (anti-)de~Sitter universe generated by null point sources with an
$m$-pole structure, localized on  the wave-front (\ref{surface}) at the singularities
$z=\pm1$, see  \cite{[B6]}. The general solution (\ref{E4.5}) corresponds
to a source described by
 \begin{equation}
  J(z,\phi) =\sum_{m=0}^\infty b_m {\epsilon(-1)^m\over8\pi} (1-z^2)^{m/2}
\left[\delta^{(m)}(z-1)+\delta^{(m)}(-z-1)\right]\cos[m(\phi-\phi_m)]\ .
\label{E2.18}
 \end{equation}
Again, each component represents a point source with an $m$-pole structure,
where the constants $b_m$ give the strength, and $\phi_m$ the
orientation of each $m$-pole.

To complete the picture, it would be natural to regard these explicit impulsive
solutions (\ref{E4.5}) in the (anti-)de~Sitter universe  generated by
{\it null} multipole particles as limits of boosted {\it static} multipole
particles, as in the case of  Minkowski background.
Unfortunately, no explicit exact solutions are known
which describe static sources of any multipole structure in a background with
$\Lambda\ne0$. Nevertheless, we were able to argue in
\cite{[B4]} that it is possible  to obtain the correct structure of the
impulsive multipole solution (\ref{general}), (\ref{E4.5}) by boosting the
perturbation of the (anti-)de~Sitter spacetime of the form (\ref{schds}) if
the function $\Psi(t,r,\vartheta)$ has an appropriate form. The corresponding
function $\H$ in (\ref{ds1}) is given by
 \begin{equation}
 \H(z,\phi) =a^3[\epsilon(1-z^2)]^{3/2} \int_0^\pi
{z^2+\cos^2\theta\over(z^2-\cos^2\theta)^2} \, {\Psi(t,r,\vartheta)\over r^2}
\,\d\theta\ ,
 \end{equation}
 in which the coordinates $t,r,\vartheta$ must be expressed using the
relations
 \begin{equation}
{\rm cosh}^2{t\over a}={z^2\sin^2\theta\over z^2-\cos^2\theta}\ ,\qquad
r={a\sqrt{\epsilon(1-z^2)}\over\sin\theta}\ , \qquad
\cos\vartheta=\sin\theta\cos\phi\ ,
 \end{equation}
 (for $\Lambda<0$ the function $\cosh$ must be replaced by $\cos$).
This is a generalization of the integral (\ref{HTimp}) which yields the
Hotta--Tanaka solution in the monopole case $\Psi=2M/r$.

\subsection{Expanding impulses generated by snapping and colliding strings}

A physically  most interesting expanding spherical impulsive
gravitational wave is probably that generated by a {\it snapping cosmic string}.
This explicit solution, described  in detail above in
section 3.5.2., can be written in various forms --- for example using
the coordinates  (\ref{GP}) with (\ref{ABsnap}), or in the form (\ref{en0}) with a
constant value $H=-{1\over2}\beta^2$ (see the transformation (\ref{trans2b}) and
the related text). Alternatively, the spacetime can also be written in the form
(\ref{en0}) with the function $H$ given by
 \begin{equation}
 H={{1\over2}\delta(1-{1\over2}\delta)\over Z^2}\ ,
 \label{H1}
 \end{equation}
where $\delta>0$. This is generated from the  function
 \begin{equation}
 h(Z)=Z^{1-\delta} ,
 \label{h1}
 \end{equation}
by the expression (\ref{Schwarz}). As shown in \cite{NutPen92}, \cite{[B10]},
the generating function $\>h\>$ is closely related to a geometrical
interpretation of the Penrose junction conditions (\ref{juncexp}).
If we evaluate the ratio $\eta/\V$ using (\ref{inv}) for $U<0$, and
using (\ref{transe3}) for $U>0$, which is related to the explicit
construction of a continuous coordinates (\ref{en0}), we observe
that on the impulse $U=0$
 \begin{equation}
{\eta\over\V}\ =\ \cases{\ Z \qquad  &for \quad $U=0_-$\ , \cr
\noalign{\medskip}
 \ h(Z) \qquad
&for \quad $U=0_+$\ . \cr }
 \label{xi}
 \end{equation}
However, it follows from (\ref{conf}), (\ref{Zcoords}) that
$\eta/\V=(x+{\hbox{i}}\,y)/(t-z)$ in Minkowski space, or similarly
$\eta/\V=(Z_2+{\hbox{i}}\,Z_3)/(Z_0-Z_1)$ in (anti-)de~Sitter space.
In both cases, this is exactly the relation for a stereographic correspondence
between a sphere and an Argand plane by  projection from the North pole
onto a plane through the equator \cite{PenRin84}. This
permits us to represent the wave surface $U=0$  either as a Riemann sphere,
or as its associated complex plane parametrized by the coordinate $Z$.
 The Penrose junction condition (\ref{juncexp})
across the wave surface then can be understood as a mapping on the complex Argand plane
$Z\to h(Z)$. According to (\ref{xi}), this is equivalent to mapping points $P_-$
on the ``inside'' of the wave surface to the identified points $P_+$ on the ``outside''.

We may  assume that the spacetime $U<0$ {\it inside} the impulse represented
by $Z=|Z|e^{i\phi}$  covers the {\it complete} sphere, $\phi\in[-\pi,\pi)$.
However, the range of the function $h(Z)$ does not cover the entire sphere
outside the spherical impulse for $U>0$. In particular, the complex mapping
(\ref{h1}) covers the plane minus a wedge as  $\>\arg h(Z)\in[-(1-\delta)\pi,(1-\delta)\pi)$.
This represents Minkowski, de Sitter, or anti-de~Sitter  space with a deficit
angle $2\pi\delta$ which may be considered to describe a snapped cosmic {\it string}
in the region {\it outside} the spherical wave. The strings have a constant tension and
are located along the axis $\eta=0$.

With the help of the above  geometrical insight we were able to construct
 even more general explicit solutions, namely  expanding impulsive waves
 generated by two {\it colliding cosmic strings}  \cite{[B10]}. In such a
situation, as first suggested by Nutku and Penrose  \cite{NutPen92},
two non-aligned  strings initially approach each other and snap at their
common point at the instant at which they collide. The remnants are four
semi-infinite strings which recede from the common point of
interaction with the speed of light, generating a specific
spherical impulse. As shown in  \cite{[B10]}, such a solution is
described  by the metric (\ref{en0}) with the structural function
of the form
 \begin{eqnarray}
 && H={{1\over2}\delta(1-{1\over2}\delta)\over Z^2}
-{{1\over2}\varepsilon(1-{1\over2}\varepsilon)16w_1^2(1-\delta)^2Z^{-2\delta}
\over
[(w_1+{\hbox{i}})Z^{1-\delta}+(w_1-{\hbox{i}})]^2
[(w_1-{\hbox{i}})Z^{1-\delta}+(w_1+{\hbox{i}})]^2}\ ,
 \label{H2c}
 \end{eqnarray}
 which is generated from
 \begin{equation}
 h(Z)=w_2{h_c^{1-\varepsilon}-1\over h_c^{1-\varepsilon}+1}
 \ ,\quad  \hbox{where}\quad
 h_c=-{(w_1-{\hbox{i}})Z^{1-\delta}+(w_1+{\hbox{i}})
 \over(w_1+{\hbox{i}})Z^{1-\delta}+(w_1-{\hbox{i}})}\  .
 \label{h2c}
 \end{equation}
Here the parameters $\delta$ and $\varepsilon$ are the deficit angles of the
first and the second pair of strings, respectively, and $w_1$, $w_2$
determine  their velocities. For more details and
illustrative pictures see \cite{[B10]}.

Let us also mention that Horta\c csu and his collaborators \cite{Hort}
analyzed vacuum fluctuations and aspects of particle creation on
the backgrounds of expanding impulsive waves.

\subsection{Geodesics in impulsive spacetimes}

In this final section we now mention some other properties
of impulsive waves in spaces of constant curvature. In particular,
we describe the behaviour of geodesics in these spacetimes.

\subsubsection{Geodesics in nonexpanding impulsive waves}

Geodesics in Minkowski space with plane-fronted impulsive {\it pp\,}-waves
were discussed in many works, e.g. in  \cite{DaT}, \cite{Sfet}, \cite{DT}.
 However, as the corresponding geodesic and geodesic deviation equations in
standard coordinates (\ref{imp-pp}) contain highly singular products of
distributions, an advanced framework of Colombeau algebras of generalized
functions had to be employed to solve these equations in
a mathematically rigorous sense \cite{Steina}.

In \cite{Marc} we have  studied the  behaviour of geodesics in spacetimes which
describe  nonexpanding impulsive waves in the (anti-)de~Sitter background.
These recent results  generalize those for impulsive {\it pp\,}-waves to the case
of a nonvanishing cosmological constant $\Lambda$. The geodesic equations  in
such spacetimes can  be derived conveniently using the  embedding of the
5-dimensional {\it pp\,}-waves (\ref{general}) onto the 4-dimensional
hyperboloid (\ref{hyp}). We obtain \cite{Marc}
\begin{eqnarray}
\ddot{ U}&=&-{\textstyle\frac{1}{3}}\Lambda\, U\,e\ ,  \nonumber \\
\ddot{ V}-{\textstyle\frac{1}{2}}\H\,\delta'( U)\,\dot{ U}^2-
 \H_{,p}\,\dot{Z_p}\,\delta( U)\,\dot{ U}
 &=& -{\textstyle\frac{1}{3}}\Lambda\, V\left[e+{\textstyle\frac{1}{2}}
  G\,\delta( U)\,\dot{ U}^2\right]\ ,  \label{5-geodeqs} \\
\ddot{Z_i}-{\textstyle\frac{1}{2}}
  \H_{,i}\,\delta( U)\,\dot{ U}^2&=&
 -{\textstyle\frac{1}{3}}\Lambda\,Z_i\left[e+{\textstyle\frac{1}{2}}
 G\,\delta( U)\,\dot{ U}^2\right]\ ,  \nonumber\\
\ddot{Z_4}-{\textstyle\frac{1}{2}}\epsilon
  \H_{,4}\,\delta( U)\,\dot{ U}^2&=&
 -{\textstyle\frac{1}{3}}\Lambda\,Z_4\!\left[e+{\textstyle\frac{1}{2}}
 G\,\delta( U)\,\dot{ U}^2\right]\ , \nonumber
 \end{eqnarray}
where $p=2,3,4$, $i=2,3$,\ $e=0,-1,+1$\  for null, timelike or
spacelike geodesics, respectively,  and
$G=Z_p\,\H_{,p}-\H$ (summation convention is
used). The coordinates \ $U\equiv{1\over\sqrt2}(Z_0+Z_1)\,$\
and \ $ V\equiv{1\over\sqrt2}(Z_0-Z_1)$\  are introduced here for
convenience, and are different from those used in the metric (\ref{conti}).
Also, we have rescaled  the structural function as
$\sqrt2\,\H\to\H$.

The equation for $ U$ in (\ref{5-geodeqs}) is clearly
decoupled and does not involve any distributional term. The solution,
which is everywhere  a {\it smooth} function of the affine parameter $\tau$,
can be written (without loss of generality) as
\begin{equation}
  U=\tau \ , \qquad
  U=a\,\dot{ U}^0\sinh(\tau/a) \ , \qquad
  U=a\,\dot{ U}^0\sin(\tau/a)\ ,  \label{U}
\end{equation}
for $\epsilon e=0$, $\epsilon e<0$  or $\epsilon e>0$, respectively.
These relations  allow us to take $\,U$ as the geodesic
parameter. Then a general solution of the remaining four functions
$Z_p$ and $V$ in (\ref{5-geodeqs}) can be written as
\begin{eqnarray}
 Z_p( U) & = & Z_p^0\,\sqrt{1-{\textstyle{1\over3}}\Lambda\,e\,{(\dot{ U}^0)}
   ^{-2}\, U^{\,2}} \ +\ (\dot Z_p^0/\dot U^0)\, U
   \ +\ A_p\,\Theta( U)\, U \ , \label{complete} \\
  V( U) & = &  V^0\,\sqrt{1-{\textstyle{1\over3}}\Lambda\,e\,{(\dot{ U}^0)}
   ^{-2}\, U^{\,2}} \ +\ (\dot V^0/\dot U^0)\, U
  \ +\ B\,\Theta( U)\,\sqrt{1-{\textstyle{1\over3}}\Lambda\,e\,{(\dot{ U}^0)}
  ^{-2}\, U^{\,2}}\nonumber\\
 &&\qquad\qquad
  \ +\ {(\dot{ U}^0)}^{-1}(\dot Z_i^0A_i+\epsilon \dot Z_4^0A_4)\,\Theta( U)\, U
  \ +\ C\,\Theta( U)\, U\ . \nonumber
\end{eqnarray}
The coefficients are
\begin{eqnarray}
 &&A_i = {\textstyle\frac{1}{2}}\left[\H_{,i}(0)
  - {\textstyle\frac{1}{3}}\Lambda\,Z_i^0G(0)\right]\ , \quad
 A_4  =  {\textstyle\frac{1}{2}}\left[\epsilon \H_{,4}(0)
  - {\textstyle\frac{1}{3}}\Lambda\,Z_4^0G(0)\right]\ , \quad
 B \,= {\textstyle\frac{1}{2}}\H(0)\ ,  \nonumber  \\
 &&C \,= {\textstyle\frac{1}{8}}\left[
   \H_{,2}^2(0)+\H_{,3}^2(0)+\epsilon \H_{,4}^2(0)+
   {\textstyle\frac{1}{3}}\Lambda\,\H^2(0)
  -{\textstyle\frac{1}{3}}\Lambda\left(Z_p^0\,\H_{,p}(0)\right)^2\right]\ ,\label{sol_BCA}
\end{eqnarray}
in which $\H(0)\equiv \H(Z_p(0))=\H(Z_p^0)$,
and the constants of integration are constrained by
\begin{eqnarray}
-2\,\dot{ U}^{\,0}\dot{ V}^{\,0}+(\dot{Z}_2^0)^2+(\dot{Z}_3^0)^2+\epsilon(\dot{Z}_4^0)^2&=&e
 \ , \nonumber\\
-2\, U^{\,0} V^{\,0}+(Z_2^0)^2+(Z_3^0)^2+\epsilon(Z_4^0)^2&=&\epsilon a^2
 \ , \label{con2}\\
- U^{\,0}\dot{ V}^{\,0}-\dot{ U}^{\,0} V^{\,0}+Z_2^0\dot{Z}_2^0+Z_3^0\dot{Z}_3^0+\epsilon Z_4^0\dot{Z}_4^0&=&0
  \ \nonumber.
\end{eqnarray}

Interestingly,  these general results also describe all geodesics in impulsive
{\it pp\,}-wave spacetimes in a Minkowski universe. For $\Lambda=0$   the geodesic
equations (\ref{5-geodeqs}) exactly reduce (omitting the coordinate $Z_4$)
to the system which has  been {\it rigorously} solved  by Kunzinger
and Steinbauer \cite{Steina} using the Colombeau algebras. Of course, the  solution
(\ref{complete}), (\ref{sol_BCA}) reduces to the explicit form of  geodesics
presented in  \cite{Steina} and  previous works
\cite{DaT}, \cite{DT}. The same results are also obtained from the
chaotic behaviour of geodesics  in nonhomogeneous sandwich
{\it pp\,}-waves  \cite{[C1]}: in the impulsive limit the chaotic motion
smears and becomes regular \cite{[C2]}.

 The main advantage of the approach presented above is that for {\it any} $\Lambda$
this yields the system (\ref{5-geodeqs}) which includes  only ``weakly'' singular
terms. In particular, there is no square of   $\delta$, contrary to other
``direct'' approaches, such as that presented in \cite{Sfet} which used the
 coordinates introduced in \cite{DaT}.

It follows from (\ref{complete}) that the geodesics are continuous but refracted by
the impulse in the transverse directions $Z_p$. However, there is a discontinuity in the
longitudinal coordinate  $V$ (and its derivative) on the impulse.
The jump on \ $ U=0$ is given by $\Delta V=B={\frac{1}{2}}\H(0)$, which is
in agreement with the Penrose junction conditions  (\ref{juncnon})
in the  ``cut and paste'' method. Further specific effects  of nonexpanding
impulses on privileged families of comoving observers associated with the natural
coordinates (\ref{globpar}) in the de~Sitter and in the anti-de~Sitter universe
were discussed in \cite{Marc} in detail. In particular, the  behaviour  of timelike
and null geodesics (including their focusing) in the  Hotta--Tanaka spacetime,
a solution with a conformally flat pure radiation impulsive wave,
and the Defrise-type impulse \cite{Defrise}  were presented.

Note finally that   symmetries of spacetimes with impulsive nonexpanding waves
were studied for $\Lambda=0$ in \cite{AiBa}, and for $\Lambda\ne0$ in \cite{Marc}.
There are always {\it at least three} Killing vector fields. In the metric form
(\ref{confppimp}), these are null rotations generated by
\begin{eqnarray}
&&x\,(\partial/\partial \V)+\U\,(\partial/\partial x)\ ,\qquad\qquad
y\,(\partial/\partial \V)+\U\,(\partial/\partial y)\ ,  \label{Killuv}\\
&&[1-{\textstyle{1\over12}\Lambda}(x^2+y^2)]\,(\partial/\partial \V)
-{\textstyle{1\over6}\Lambda}\,\U\left[ x\,(\partial/\partial x)
+y\,(\partial/\partial y)+\U\,(\partial/\partial \U)\right] \ , \nonumber
\end{eqnarray}
where $\eta={1\over\sqrt2}(x+\hbox{i}\,y)$. The Killing vectors (\ref{Killuv})
exactly reduce to those found previously by
Aichelburg and Balasin \cite{AiBa} for impulsive {\it pp\,}-waves
in the Minkowski background. Additional symmetries arise for  specific forms
of the structural function. For example, in  case of the Hotta--Tanaka
solution (\ref{HTimp}) the function $\H$ only depends on
$z=\sqrt{1-\epsilon a^{-2}(Z_2^2+Z_3^2)}$. There is thus also an
axial symmetry. Again, this is analogous to the axisymmetric Aichelburg--Sexl
solution for the case $\Lambda=0$ which also admits four Killing vectors \cite{AiBa}.

\subsubsection{Geodesics in expanding impulsive waves}

As far as we know, an explicit form for  geodesics in spacetimes with
expanding spherical gravitational impulses has not  been presented in
the literature so far. It is possible to derive such geodesics assuming
these are  $C^1$ in the continuous coordinate system (\ref{en0}).
With this assumption, the constants
\begin{eqnarray}
&&Z_i\equiv Z(\tau_i)\ ,\quad V_i\equiv V(\tau_i)\ ,\quad
  U_i\equiv U(\tau_i)=0\ ,\nonumber\\
&&\dot Z_i\equiv \dot Z(\tau_i)\ ,\quad\dot V_i\equiv \dot V(\tau_i)\ ,\quad
  \dot U_i\equiv \dot U(\tau_i)\ ,
\label{const}
\end{eqnarray}
which characterize the  positions and velocities  at $\tau_i$, the instant of
interaction with the impulsive wave, must have the same values when evaluated
in the limits $U\to0$ both from the region in front $(U>0)$ and behind $(U<0)$
the impulse. Let us consider here only the case $\Lambda=0$ for the sake of
simplicity. We start with a general free motion
\begin{eqnarray}
&&t^-=\gamma\,\tau\ ,  \nonumber\\
&&x^-=\dot x_0\,(\tau-\tau_i)+x_0\ , \label{geod+b}\\
&&y^-=\dot y_0\,(\tau-\tau_i)+y_0\ , \nonumber\\
&&z^-=\,\dot z_0\,(\tau-\tau_i)+z_0\ , \nonumber
\end{eqnarray}
where
$\gamma=\sqrt{\dot x_0^2+\dot y_0^2+\dot z_0^2-e}$,
in the flat Minkowski region $U<0$. It is straightforward to express the constants
(\ref{const}) in terms of the initial data (\ref{geod+b}) using the inverse to
(\ref{inv}) and standard relations
${\cal U}=\textstyle{\frac{1}{\sqrt2}}(t+z)$,
${\cal V}=\textstyle{\frac{1}{\sqrt2}}(t-z)$,
$\eta=\textstyle{\frac{1}{\sqrt2}}(x+\,\hbox{i}\,y)$,
as
\begin{eqnarray}
&&Z_i={x_0 +\, \hbox{i}\, y_0\over \gamma\tau_i-z_0}\ ,\qquad
V_i={\textstyle{1\over\sqrt2}}\,[(1+\epsilon)\gamma\tau_i-(1-\epsilon)
z_0]\ ,\nonumber\\
&&\dot Z_i={\dot x_0 +\, \hbox{i}\, \dot y_0\over \gamma\tau_i-z_0}
-{x_0 +\, \hbox{i}\, y_0\over (\gamma\tau_i-z_0)^2}\nonumber\\
&&\hskip15mm \times{[(1-\epsilon)\gamma-(1+\epsilon)\dot z_0](\gamma\tau_i-z_0)
  +2\epsilon(x_0\dot x_0+y_0\dot y_0)
\over {(1+\epsilon) \gamma\tau_i-(1-\epsilon)z_0}} \ ,\label{genercont}\\
&& \dot V_i={\textstyle{1\over\sqrt2}}\,\Big\{
[(1-\epsilon)\gamma- (1+\epsilon)\dot z_0]\,
[(1-\epsilon)\gamma\tau_i-(1+\epsilon)z_0]  \nonumber\\
&&\hskip15mm +4\epsilon(x_0\dot x_0+y_0\dot y_0)\Big\}\,
  [(1+\epsilon) \gamma\tau_i-(1-\epsilon)z_0]^{-1}\ ,\nonumber\\
&&\dot U_i={\textstyle{\sqrt2}}\,\,
{x_0\dot x_0+y_0\dot y_0+z_0\dot z_0-\gamma^2\tau_i\over
(1+\epsilon)\gamma\tau_i-(1-\epsilon) z_0}\ .\nonumber
\end{eqnarray}
Now, we express the  geodesics outside the impulse
in the (locally) Minkowski space $U>0$
\begin{eqnarray}
{\cal V}^+ &=&\dot{\cal V}_0^+(\tau-\tau_i)+{\cal V}_0^+\ ,\nonumber\\
{\cal U}^+ &=&\dot{\cal U}_0^+(\tau-\tau_i)+{\cal U}_0^+\ ,\label{geod++}\\
\eta^+&=&\dot\eta_0^+(\tau-\tau_i)+\eta_0^+\ ,\nonumber
\end{eqnarray}
using (\ref{transe3}) as
\begin{eqnarray}
&& {\cal V}_0^+=AV_i\ ,\qquad
{\cal U}_0^+=BV_i\ ,\qquad
\eta_0^+\,=\,CV_i\ ,\qquad \nonumber\\
&&\dot{\cal V}_0^+=A\dot V_i-D \dot U_i
   +(A_{,Z}\dot Z_i+A_{,\bar Z}\dot{\bar Z}_i) V_i\ ,\label{complcoef++}\\
&&\dot{\cal U}_0^+=B\dot V_i-E \dot U_i
   +(B_{,Z}\dot Z_i+B_{,\bar Z}\dot{\bar Z}_i) V_i\ , \nonumber\\
&&\dot\eta_0^+\,=C\dot V_i-F \dot U_i
   +(C_{,Z}\dot Z_i+C_{,\bar Z}\dot{\bar Z}_i) V_i \ ,\nonumber
\end{eqnarray}
in which the coefficients  and their derivatives are given by the
functions (\ref{transe4}) evaluated at $Z_i$.

These general results can be used for a physical discussion of geodesics in specific impulsive
solutions  \cite{Roland}, such as the expanding spherical impulse
generated by a snapping cosmic string which we have described above in sections 3.5.2 and 4.2.

\section{Repetition and  a few final remarks}

In this essay, we have presented a brief review of exact solutions of  Einstein's
equations which describe impulsive  waves in spaces of constant curvature.
These are either gravitational and/or null matter nonexpanding impulses, or
expanding spherical impulsive purely gravitational waves (attached to cosmic
strings) which propagate in Minkowski, de~Sitter, or anti-de~Sitter universes.
We have systematically and explicitly described  all the main methods for
their construction: the  ``cut and paste'' method,  introduction of the continuous
coordinate system, considering distributional limits of sandwich waves, geometrical
embedding from higher dimensions,  and boosts or limits of infinite
acceleration of some initially static or accelerating sources.
In the further part of the contribution we concentrated on some
of their properties, in particular on the behaviour of geodesics.
Physically interesting special solutions were also emphasized.

We have tried to provide an overall review. Nevertheless, we are aware of
the fact that the point of view presented could depend on our own approach
to the subject. The essay may thus be biased, as more space was certainly
devoted to a description of those aspects of impulsive spacetimes which we had
personally  investigated. However, we attempted to present a comprehensive
list of the relevant references. If some are missing this is
unintentional, and the author deeply apologizes for this.

We have dedicated the  essay to Professor Bi\v c\'ak on the occasion of
the landmark anniversary of his birthday. It has been demonstrated that he
 contributed significantly also to the investigation of impulsive
waves. Of course, this is just one small part of his both
extensive and intensive work in the field of exact radiative
spacetimes.

Let me finally close this contribution by personal words once more.
It was Professor Bi\v c\' ak who put me wise to many of the secrets,
mysteries, and wonders of Einstein's theory. He has also led my first
steps in the world of relativity, formed and guided my professional
activities. From the very beginning I have always admired his
approach to general relativity. This is based on a deep understanding
of the geometry and global structure of investigated spacetimes, yet with the
emphasis being placed on presenting their clear physical interpretation.
I hope that I have succeeded to emulate (at least partially) his ``style''
here.  And I also hope that
Ji\v r\'\i\  Bi\v c\' ak will not take his pencil during the
reading of this essay, making many strokes, remarks, corrections,
amendments, and improvements, as he has done so many times before
with  my previous manuscripts.

\section*{Acknowledgements}

I wish to thank very much Jerry Griffiths, my long-standing collaborator on
impulsive waves, for all his suggestions and comments concerning this review.
Special thanks are due to my wife for her patience and understanding.
The work was supported by  the grant GACR-202/99/0261
from the Czech Republic, and GAUK 141/2000 of Charles University in Prague.

\end{document}